\documentclass[11pt, a4paper]{article}
\pdfoutput=1
\usepackage{jcappub}

\usepackage{booktabs}
\usepackage{multirow}
\usepackage{amssymb}
\usepackage[export]{adjustbox}
\usepackage{floatrow}
\newfloatcommand{capbtabbox}{table}[][3.5in]
\newfloatcommand{capbfigbox}{figure}[][2.3in]
\usepackage{blindtext}
\usepackage{amsmath}
\usepackage{booktabs}
\usepackage{aas_macros}
\usepackage{soul}
\usepackage[normalem]{ulem}

\newcommand{\ie}{i.e.~}
\newcommand{\eg}{e.g.~}

\def\lsim{\mathrel{\raise.3ex\hbox{$<$\kern-.75em\lower1ex\hbox{$\sim$}}}}
\def\gsim{\mathrel{\raise.3ex\hbox{$>$\kern-.75em\lower1ex\hbox{$\sim$}}}}

\usepackage{appendix}
\usepackage{array}
\newcolumntype{x}[1]{>{\centering\arraybackslash}p{#1}}
\usepackage{bm}
\usepackage{color}
\usepackage{graphicx}
\usepackage{cancel}
\usepackage{amsfonts}
\usepackage{amssymb}
\usepackage{amsmath}
\usepackage{calc}
\usepackage{dcolumn}
\usepackage{mathrsfs}
\usepackage{mathtools}

\newcommand{\Eq}[1]{Eq.~\eqref{#1}}
\newcommand{\Fig}[1]{Fig.~\ref{#1}}
\newcommand{\Sec}[1]{Sec.~\ref{#1}}

\newcommand{\Ref}[1]{Ref.~\cite{#1}}     

\newcommand{\beq}{\begin{equation}}
\newcommand{\eeq}{\end{equation}}

\definecolor{rossoCP3}{cmyk}{0,.88,.77,.40}
\definecolor{verdeCP3}{rgb}{0.09765625, 0.57421875, 0.1015625}
\definecolor{bluCP3}{rgb}{0, 0.23, 0.67}

\ifx\pdfoutput\undefined
\usepackage[dvips,bookmarks]{hyperref}    
\else
\usepackage{hyperref}    
\fi
\hypersetup{colorlinks, bookmarksopen, bookmarksnumbered,
citecolor=verdeCP3, linkcolor=bluCP3, pdfstartview=FitH, urlcolor=rossoCP3}

\begin{document}

\hspace*{110mm}{\large \tt FERMILAB-PUB-16-470-A}

\vskip 0.2in

\title{Gamma Rays From Dark Matter Subhalos Revisited: Refining the Predictions and Constraints}

\author[a,b,c]{Dan Hooper}\note{ORCID: http://orcid.org/0000-0001-8837-4127}
\emailAdd{dhooper@fnal.gov}
\author[a,d]{and Samuel J.~Witte}\note{ORCID: http://orcid.org/0000-0003-4649-3085}
\emailAdd{switte@physics.ucla.edu}

\affiliation[a]{Fermi National Accelerator Laboratory, Center for Particle
Astrophysics, Batavia, IL 60510}
\affiliation[b]{University of Chicago, Department of Astronomy and Astrophysics, Chicago, IL 60637}
\affiliation[c]{University of Chicago, Kavli Institute for Cosmological Physics, Chicago, IL 60637}
\affiliation[d]{University of California, Los Angeles, Department of Physics and Astronomy, Los Angeles, CA 90095}

\abstract{Utilizing data from the ELVIS and Via Lactea-II simulations, we characterize the local dark matter subhalo population, and use this information to refine the predictions for the gamma-ray fluxes arising from annihilating dark matter in this class of objects. We find that the shapes of nearby subhalos are significantly altered by tidal effects, and are generally not well described by NFW density profiles, instead prefering power-law profiles with an exponential cutoff. From the subhalo candidates detected by the Fermi Gamma-Ray Space Telescope, we place limits on the dark matter annihilation cross section that are only modestly weaker than those based on observations of dwarf galaxies. We also calculate the fraction of observable subhalos that are predicted to be spatially extended at a level potentially discernible to Fermi.}

\maketitle

\section{Introduction}

A wide range of experimental strategies are being pursued in an effort to observe dark matter's non-gravitational interactions and ultimately identify the particle nature of dark matter. This program includes but is not limited to experiments designed to detect the scattering of dark matter with nuclei, searches for the annihilation or decay products of dark matter, and efforts to produce and observe dark matter at accelerators. In all three of these areas, current experiments are probing important regions of parameter space and are sensitive to a wide range of well motivated dark matter candidates.

Particularly promising are searches for dark matter utilizing gamma-ray telescopes. Constraints from the Fermi Gamma-Ray Space Telescope's observations of dwarf spheroidal galaxies~\cite{Drlica-Wagner:2015xua,Geringer-Sameth:2014qqa}, the Galactic Center~\cite{Hooper:2012sr} and the extragalactic gamma-ray background~\cite{Ackermann:2015tah,DiMauro:2015tfa}, are each currently sensitive to dark matter particles with masses in the range of $\sim$10-100 GeV and annihilation cross sections similar to that naively predicted from thermal relic abundance considerations, $\sigma v \simeq 2 \times 10^{-26}$ cm$^3/$s. Furthermore, the excess of GeV-scale gamma rays observed from the region surrounding the Galactic Center exhibits a spectrum and morphology that is consistent with the predictions of dark matter particles with a mass of $\sim 30-60$ GeV and an annihilation cross section of $\sigma v \sim 10^{-26}$ cm$^3/$s~\cite{TheFermi-LAT:2015kwa,Daylan:2014rsa,Calore:2014xka,Abazajian:2014fta,Hooper:2013rwa,Gordon:2013vta,Abazajian:2012pn,Hooper:2011ti,Hooper:2010mq,Goodenough:2009gk} (for discussions of other interpretations, see Refs.~\cite{Cholis:2014lta,Lee:2015fea,Bartels:2015aea,Petrovic:2014xra,Hooper:2013nhl,Hooper:2016rap,Brandt:2015ula,Petrovic:2014uda,Cholis:2015dea,Carlson:2014cwa}).

Within the standard paradigm of cold and collisionless dark matter, structure forms hierarchically, meaning that the smallest halos form first and gradually merge to form larger systems, including the halos that host galaxies and galaxy clusters~\cite{White:1991mr}. As a consequence of this process, the dark matter halos that encompass galaxies are predicted to contain large numbers of smaller subhalos. 


In the case of the Milky Way, the largest and most massive subhalos include the known dwarf galaxies, as well as the Large and Small Magellanic Clouds. This collection of very massive objects reflects only a small fraction of the subhalo population, however.  A much larger number of subhalos that are too small to capture significant quantities of gas and form stars are also expected to be present, while remaining invisible to surveys at optical and other wavelengths. If dark matter particles annihilate with a cross section that is similar to that naively predicted for a thermal relic, nearby subhalos could be a promising target for gamma-ray telescopes~\cite{Bertoni:2015mla,Bertoni:2016hoh,Schoonenberg:2016aml,Berlin:2013dva,Belikov:2011pu,Buckley:2010vg,Ackermann:2012nb,Zechlin:2011wa,Mirabal:2012em,Mirabal:2010ny,Zechlin:2012by,Kuhlen:2008aw,Pieri:2007ir,Baltz:2008wd,Springel:2008by,Springel:2008zz,Koushiappas:2003bn,Tasitsiomi:2002vh,Ishiyama:2010es,Hutten:2016jko}.

The most recent catalog released by the Fermi Collaboration (the 3FGL) contains 992 gamma-ray sources that have not been associated with emission observed at other wavelengths~\cite{TheFermi-LAT:2015hja}, a small fraction of which could potentially be dark matter subhalos. Recent studies of the 3FGL identified a subset of 19 bright ($\Phi_{\gamma}>7 \times 10^{-10}$ cm$^{-2}$ s$^{-1}$, $E_{\gamma} > 1$ GeV) and high-latitude ($|b|>20^{\circ}$) sources that show no evidence of variability and exhibit a spectral shape that is consistent with the predictions of annihilating dark matter~\cite{Bertoni:2015mla,Bertoni:2016hoh}. From the characteristics of these subhalo candidate sources, limits can be derived on the dark matter annihilation cross section. Such limits, however, can vary significantly depending on the assumptions that are made regarding the local abundance of dark matter subhalos and on the spatial distribution of dark matter within these systems. For example, the limits on the dark matter's annihilation cross section placed in Ref.~\cite{Bertoni:2015mla} and Ref.~\cite{Schoonenberg:2016aml} differ by a factor of a few for most dark matter masses. Actually, there are a number of significant differences between the analyses of Ref.~\cite{Bertoni:2015mla} and Ref.~\cite{Schoonenberg:2016aml} which mitigate their apparent disagreement. Specifically, the mass range analyzed by Ref.~\cite{Schoonenberg:2016aml} extends 2-3 orders of magnitude above what was used in Ref.~\cite{Bertoni:2015mla}, and the density profiles and halo-to-halo variations used in Ref.~\cite{Bertoni:2015mla} predict significantly higher gamma-ray fluxes for the same mass subhalos. The various assumptions entering each of these analyses seem at face value to be quite reasonable. Consider, for example, the density profiles used to characterize the local subhalo population. The authors of \Ref{Bertoni:2015mla} adopted density profiles that are described by an Einasto profile, tidally truncated to remove the outermost 99.5\% of a given subhalo's mass, \Ref{Schoonenberg:2016aml} chose instead to adopt a traditional NFW density profile, with concentrations chosen to match the parameters of a given subhalo identified within the Via Lactea II simulation. In reality, however, it is likely that the true population of nearby subhalos is not particularly well described by either of these simple halo profile parameterizations. 

Ref.~\cite{Hutten:2016jko} has also recently performed a more comprehensive assessment of how various uncertainties associated with the subhalo distribution and parameterization may effect their observability with gamma-ray telescopes. In light of the large variability that is produced from seemingly reasonable assumptions, it has become clear that a focused and self-consistent analysis of the local subhalo population is necessary before reliable statements can be made regarding subhalo detectability.

In this paper, we revisit the characteristics of the local dark matter subhalo population, basing our analysis on the properties of the subhalos identified within the cosmological simulations Via Lactea II and ELVIS. We find that the simulated subhalos in the local region of the Milky Way are generally well characterized by power-law density profiles with an exponential cutoff. Using this profile parameterization, and accounting for halo-to-halo variations as determined by the distribution of simulated subhalos, we estimate the number of subhalos that could be observed by the Fermi Large Area Telescope (Fermi-LAT) and use this information to place constraints on the dark matter's annihilation cross section. We also calculate the fraction of the observable subhalos that will be spatially extended at a level potentially discernible to experiments such as Fermi, providing us with a way of discriminating a dark matter subhalo population from a collection of point-like gamma-ray sources.


\newpage

\section{Subhalo Populations in Cosmological Simulations \label{sec:cosmosims}}

Various groups have utilized Fermi's catalog of unassociated gamma-ray sources to derive limits on the dark matter annihilation cross section~\cite{Berlin:2013dva, Bertoni:2015mla, Schoonenberg:2016aml, Mirabal:2016huj}. The results of these studies, however, vary considerably depending on the assumed characteristics of the local subhalo population. Among the least understood characteristics, is the response of the subhalo density profile to extreme tidal forces. 

There have been numerous attempts to study the intimate details of the subhalo radial density profile (\eg see~\cite{Moore:1997sg,Ghigna:1999sn,Moore:1999gc,Hayashi:2002qv,Kazantzidis:2003hb,Taffoni:2003sr,Taylor:2003ja,RomanoDiaz:2006ep,Sales:2007hp,Diemand:2007qr,Penarrubia:2007zx,Choi:2008xi,Penarrubia:2010jk,Ishiyama:2010es,Oh:2010mc,Ishiyama:2014uoa,Jiang:2014nsa,Bosch:2014qsa,Oh:2015xoa,Moline:2016pbm} for an incomplete list). Specifically, these studies have largely focused on using either high-resolution simulations or semi-analytic tools to study the process by which tidal forces of the host halo disrupt the subhalo's density distribution. By analyzing the distribution of test particles within the tidally disrupted subhalos, various groups (see \eg \cite{Hayashi:2002qv,Kazantzidis:2003hb,Penarrubia:2010jk}) have derived modified profiles, often taken to be extensions of the canonical NFW profile, that characterize the resolved subhalo profile as a function of \eg mass, location, orbit, merger history, etc. 

More recently, attempts have been made to simplify the characterization of these subhalos for the more practical purpose of implementing these modifications into calculations. For example, Ref.~\cite{Moline:2016pbm} attempted to characterize the subhalo population identified in the Via Lactea II and ELVIS simulations using an NFW profile, but with concentration parameters that were dependent on both the subhalo mass  (or maximum circular velocity) and the location of the subhalo relative to the host center. This was done for the purpose of calculating boost factors. 

Here in \Sec{sec:cosmosims}, we take a similar approach to \cite{Moline:2016pbm}, in that we adapt a more generalized parameterization of subhalos identified in the Via Lactea II and ELVIS simulations for the purpose of assessing the impact that tidal stripping has on the observability of subhalos. The primary difference between our approach to characterizing these subhalos and that of \cite{Moline:2016pbm}, is that we relax the assumption that tidally stripped halos are well-described by an NFW profile, and instead attempt to parametrize density distributions with a mass and location dependent profile. Thus, by deriving subhalo characteristics and distributions from a fixed set of simulations, we attempt here to provide a more self-consistent and reliable description of the observability of dark matter subhalos.

\subsection{The Via Lactea II and ELVIS Simulations \label{sec:vl2el}}

In an effort to characterize the population of dark matter subhalos located within the local volume of the Milky Way, we utilize the publicly available data from the Via Lactea II (VL-II)~\cite{Diemand:2008in} and ELVIS~\cite{Garrison-Kimmel:2013eoa} cosmological simulations. The VL-II simulation contains over 1 billion particles, each with a mass of $4.1 \times 10^3 M_\odot$, and identifies approximately $20,000$ subhalos with a maximum circular velocity, $v_{c,{\rm max}}$, greater than $4$ km/s. Since we are interested here in subhalos residing within Milky Way-like halos, we have chosen to restrict our attention to those subhalos that are located within $300$ kiloparsecs (kpc) of the center of the host halo. Furthermore, in order to minimize the impact of thresholds and other ambiguities associated with subhalo identification and characterization, we limit our analysis to those subhalos that consist of $100$ or more particles. These cuts reduce the number of VL-II subhalos used in our analysis to $5,268$.

The ELVIS suite consists of 48 simulated halos, each comprised of at least $53$ million particles with masses of $1.9 \times 10^5 M_\odot$. Half of these simulations are of paired galaxies, intended to be representative of the Milky Way-Andromeda system in both mass and phase space. Three high-resolution simulations were performed on isolated halos (in addition to the aforementioned 48) with a particle mass of $2.35 \times 10^4 M_\odot$. In our analysis, we consider those subhalos that are comprised of at least $100$ particles and with $v_{c,{\rm max}} > 8$ km/s in the paired and isolated simulations, and $v_{c,{\rm max}} > 4$ km/s in the high-resolution simulations. As with VL-II, we have restricted our attention to subhalos that are located within $300$ kpc of the nearest host halo's center, leaving us with a total of $26,048$ subhalos from among the suite of ELVIS simulations. 

For each subhalo found in either simulation catalogue, we extract $v_{c,{\rm max}}$, the radius at which maximum circular velocity occurs, $R_{v,{\rm max}}$, and the total gravitationally bound mass (each evaluated at $z=0$). It is well known that the velocity profiles and concentrations of the subhalos extracted from dark matter simulations depend on the precise values of the adopted cosmological parameters. \Ref{Polisensky:2013ppa} derived a scaling relation for $R_{v,{\rm max}}$ (at fixed $v_{c,{\rm max}}$) on the cosmological parameters $\sigma_8$ and $n_s$, based on the results of various cold DM simulations. Specifically, they found the following:
\begin{equation}
R_{v,{\rm max}} \propto (\sigma_8 \, 5.5^{n_s})^{-1.5} \, .
\end{equation}
Since the cosmological parameters adopted by VL-II and ELVIS are based on WMAP-3 ($\sigma_8=0.74$, $n_s=0.951$) and WMAP-7 ($\sigma_8=0.80$, $n_s=0.963$), respectively, we have rescaled both to the latest results from the Planck Collaboration ($\sigma_8 = 0.82$, $n_s=0.967$)~\cite{Ade:2015xua}.

\subsection{The Dark Matter Profiles of Simulated Subhalos \label{sec:subparamet}}

Here, we investigate the distribution of dark matter in subhalos identified within the VL-II and ELVIS simulations. Specifically, for each subhalo, we considered various parameterizations of the density profile and determined which can provide good agreement with the simulated values of $v_{c,{\rm max}}$, $R_{v,{\rm max}}$, and the total gravitationally bound mass.\footnote{Although we would ideally like to extract information for the $r < R_{v,\text{max}}$ region of a given subhalo, statistical limitations make this impractical in most cases. We focus here on the values of the more reliably determined quantities, $v_{c,{\rm max}}$, $R_{v,{\rm max}}$, and $M_{\rm tot}$.} 

After determining that the subhalo profile parameterizations adopted in both Refs.~\cite{Berlin:2013dva, Bertoni:2015mla} and Ref.~\cite{Schoonenberg:2016aml} provide poor fits to the subhalos located near the center of the host halo, we further considered a doubly-generalized NFW profile of the following form:
\begin{equation}
\rho(r) = \frac{\rho_s}{\left(\frac{r}{r_s}\right)^{\gamma_{_1}}\left(\frac{r}{r_s} +1\right)^{\gamma_{_2}}},
\end{equation}
where the case of a canonical NFW profile is recovered for $\gamma_{_1}=1$ and $\gamma_{_2}=2$. For those subhalos located in the outer regions of a host halo, we found that this parameterization could in most cases be tuned to match the characteristics found in the simulations. But for subhalos located within the innermost few tens of kiloparsecs of their host halo, we found that this class of profile shapes could generally {\it not} simultaneously accommodate both the mass contained within $R_{v,\text{max}}$ (i.e. $M(<R_{v,\text{max}})$) and the total mass, $M_\text{tot}$, of the subhalo (for any profile with $\gamma_{_1}>0$). We attribute the inability of the doubly-generalized NFW profile to describe these subhalos to the effects of tidal stripping, which are more pronounced in high density environments.

Next, inspired by \Ref{Kazantzidis:2003hb} (see also Ref.~\cite{Penarrubia:2007zx}), we considered the following density profile for the local population of tidally truncated subhalos:
\begin{equation}\label{eq:den_trunc}
\rho(r) = \frac{\rho_0}{r^{\gamma}} \, \exp\bigg(-\frac{r}{R_b}\bigg) \, .
\end{equation}
For nearly all of the simulated subhalos considered in this analysis, we are able to identify choices of $\gamma$ and $R_b$ that can simultaneously accommodate the reported values of both $M(<R_{v,{\rm max}})$ and $M_{\rm tot}$.

\begin{figure*}
\center
\includegraphics[width=.49\textwidth, trim=0.4cm 0.0cm 0.4cm 0.4cm,clip=true]{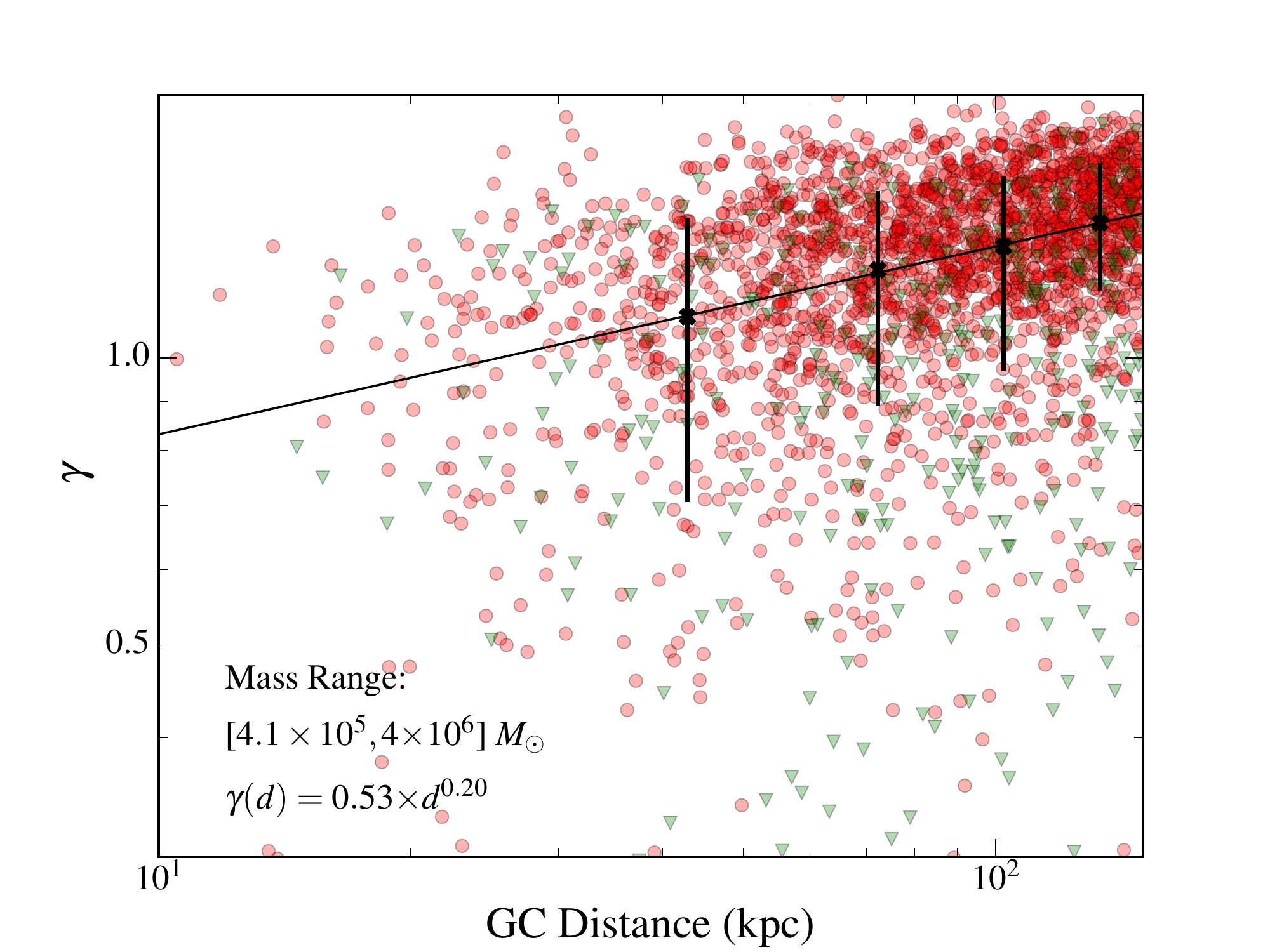}
\includegraphics[width=.49\textwidth, trim=0.4cm 0.0cm 0.4cm 0.4cm,clip=true]{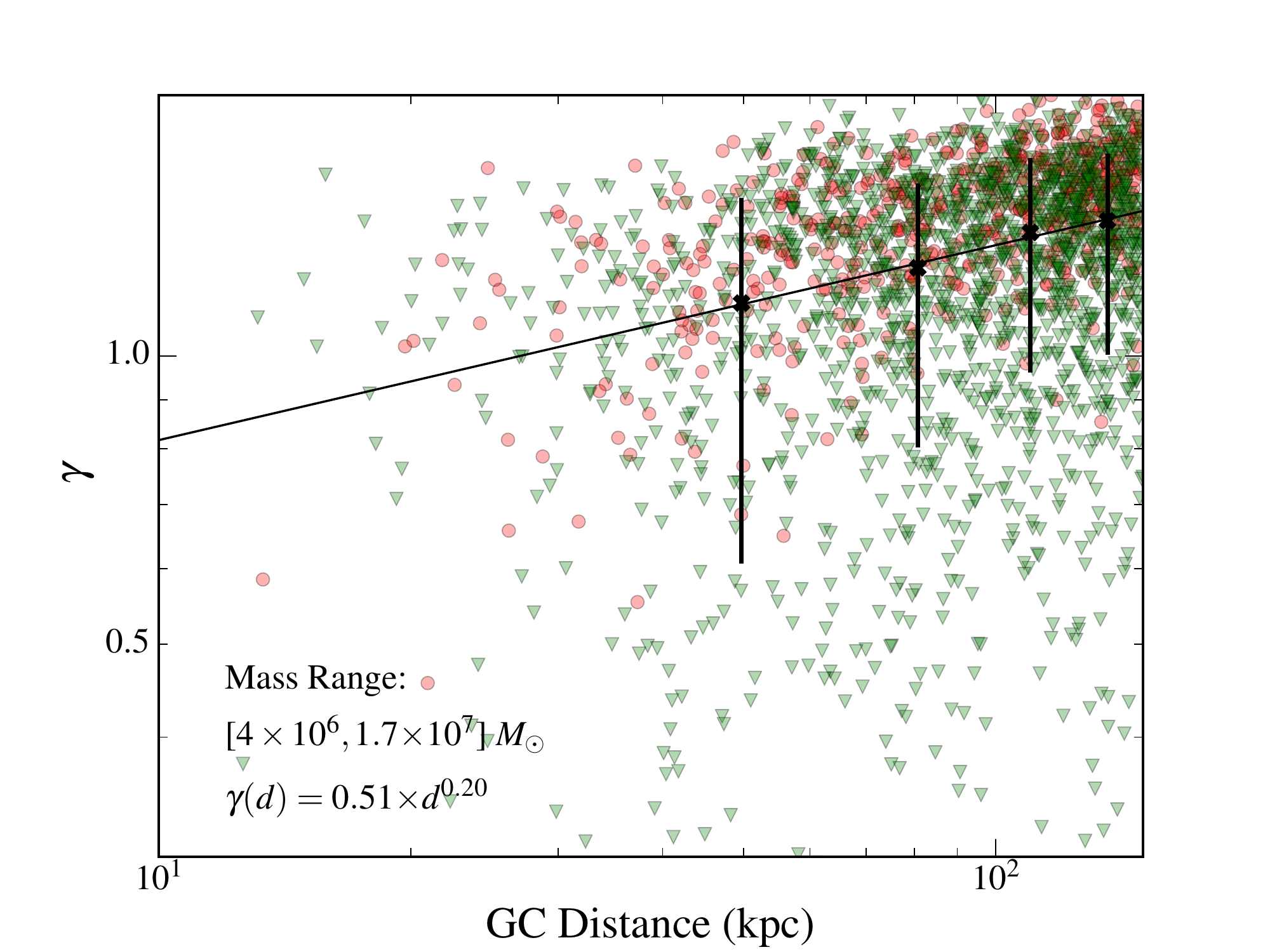}
\includegraphics[width=.49\textwidth,trim=0.4cm 0.0cm 0.4cm 0.4cm,clip=true]{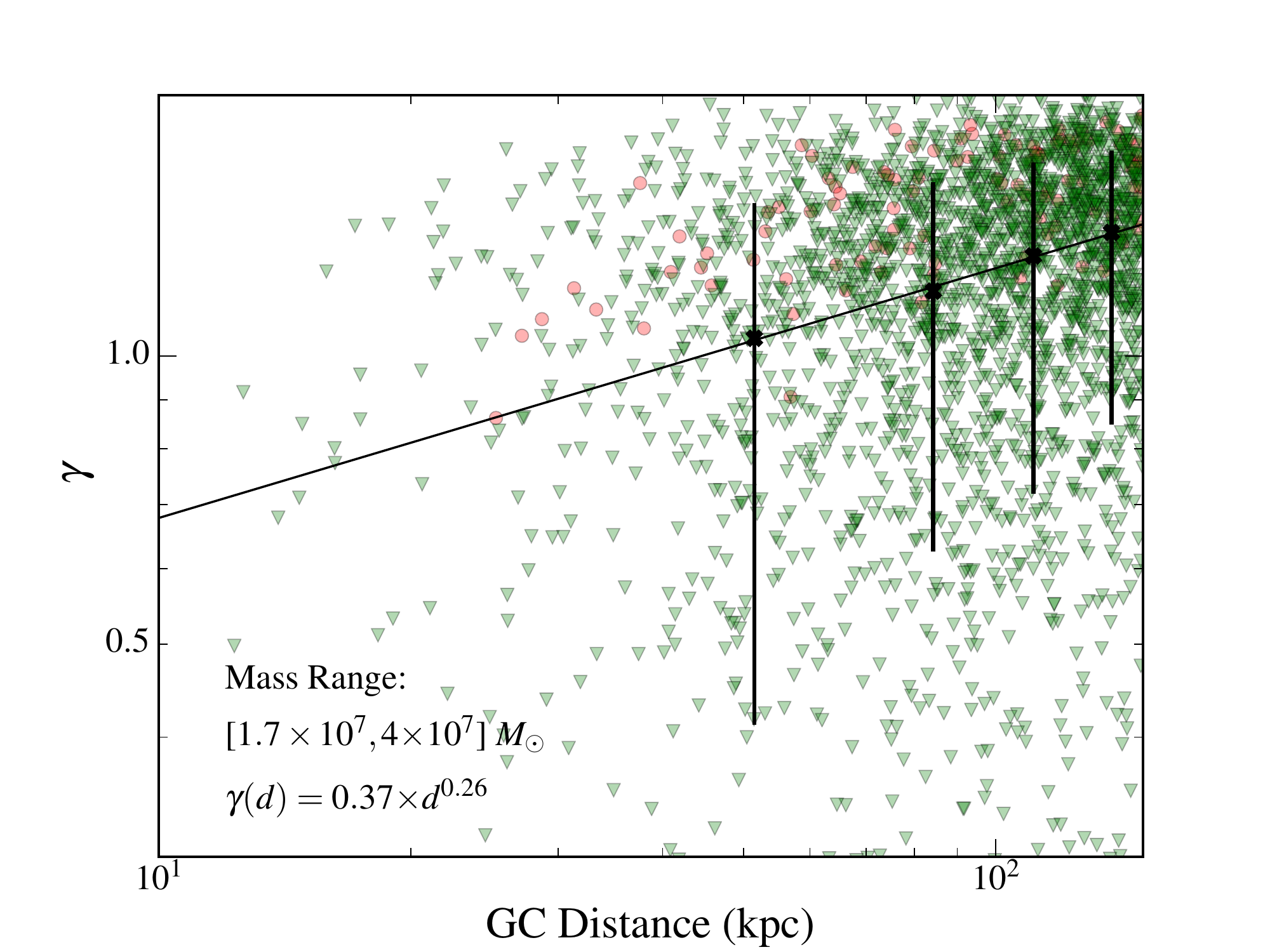}
\includegraphics[width=.49\textwidth, trim=0.4cm 0.0cm 0.4cm 0.4cm,clip=true]{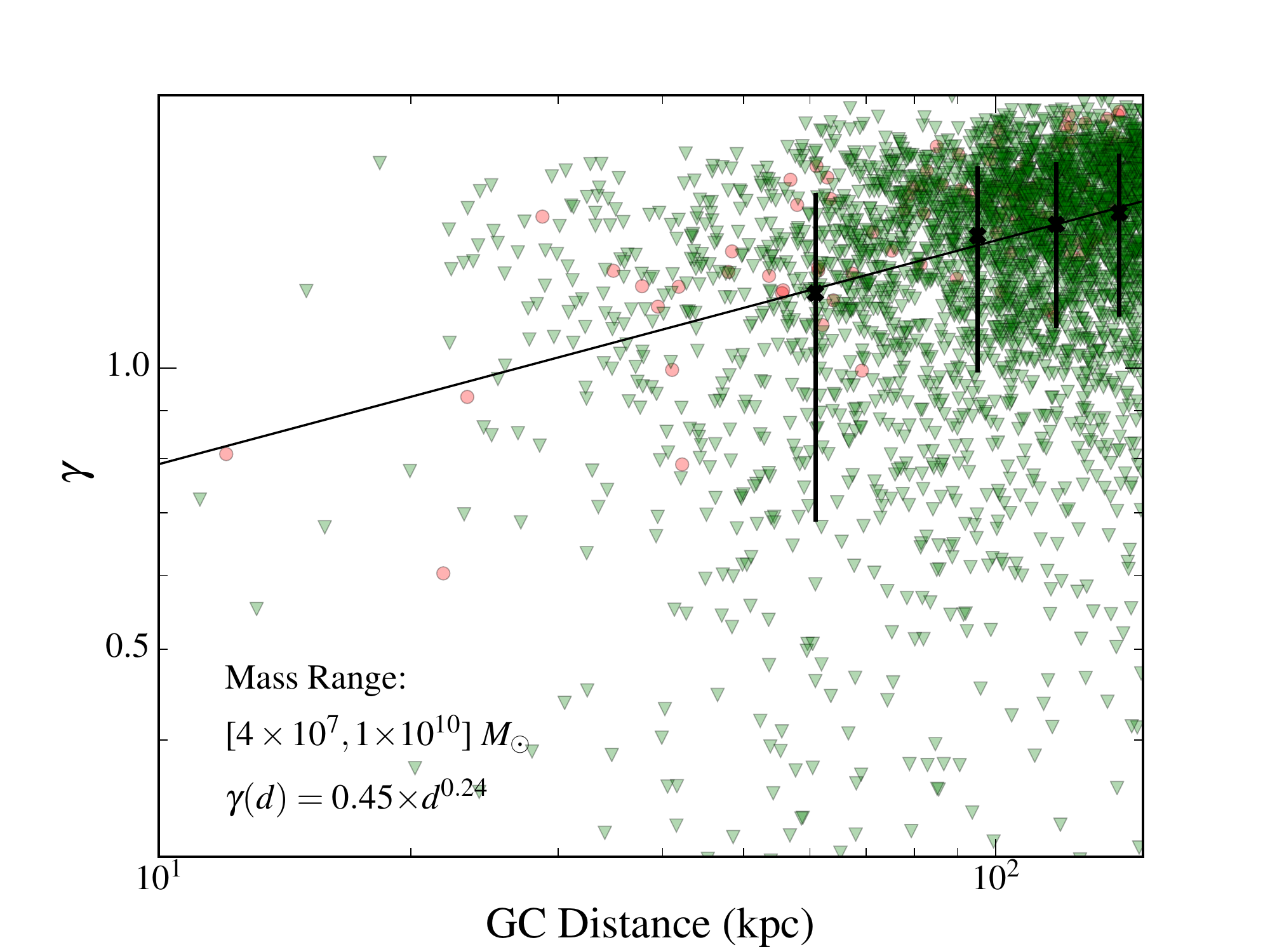}
\caption{\label{fig:scatter1} The best-fit values found for the inner slope, $\gamma$, for subhalos in the Via Lactea-II (red circles) and ELVIS simulations (green triangles), adopting a profile of the form $\rho(r) =\rho_0 \, r^{-\gamma} \exp(-r/R_b)$. Results are presented as a function of the distance of the subhalo to the center of the host halo, with each frame corresponding to subhalos in a different mass range.  The solid line denotes the power-law trend for the median value of this parameter (the equations for which are given in each panel, denoted $\gamma(d)$), while the error bars depict the range of values found among the central 68\% of subhalos in each of four distance bins.  }
\end{figure*}

\begin{figure*}
\center
\includegraphics[width=.49\textwidth, trim=0.4cm 0.0cm 0.4cm 0.4cm,clip=true]{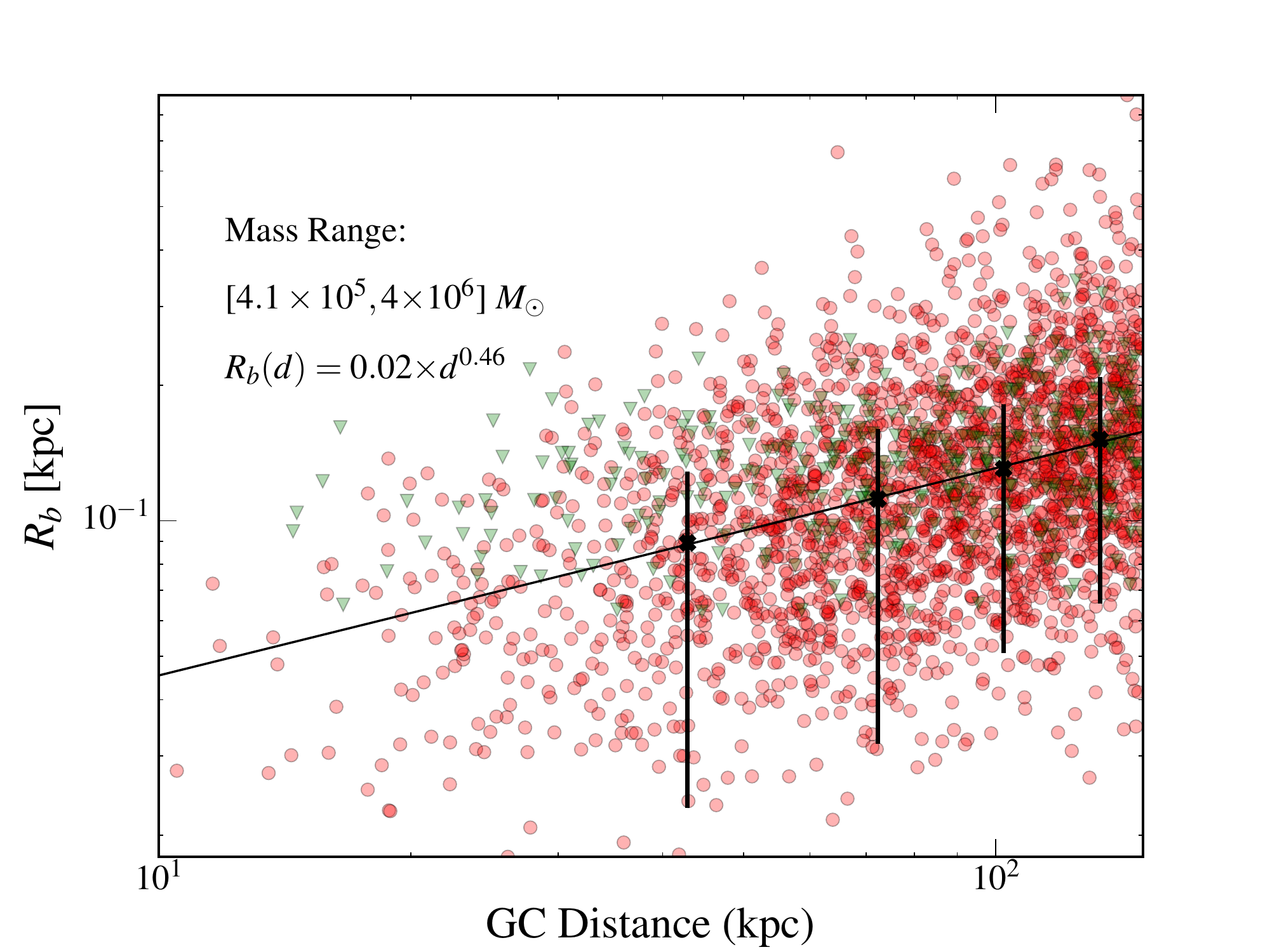}
\includegraphics[width=.49\textwidth, trim=0.4cm 0.0cm 0.4cm 0.4cm,clip=true]{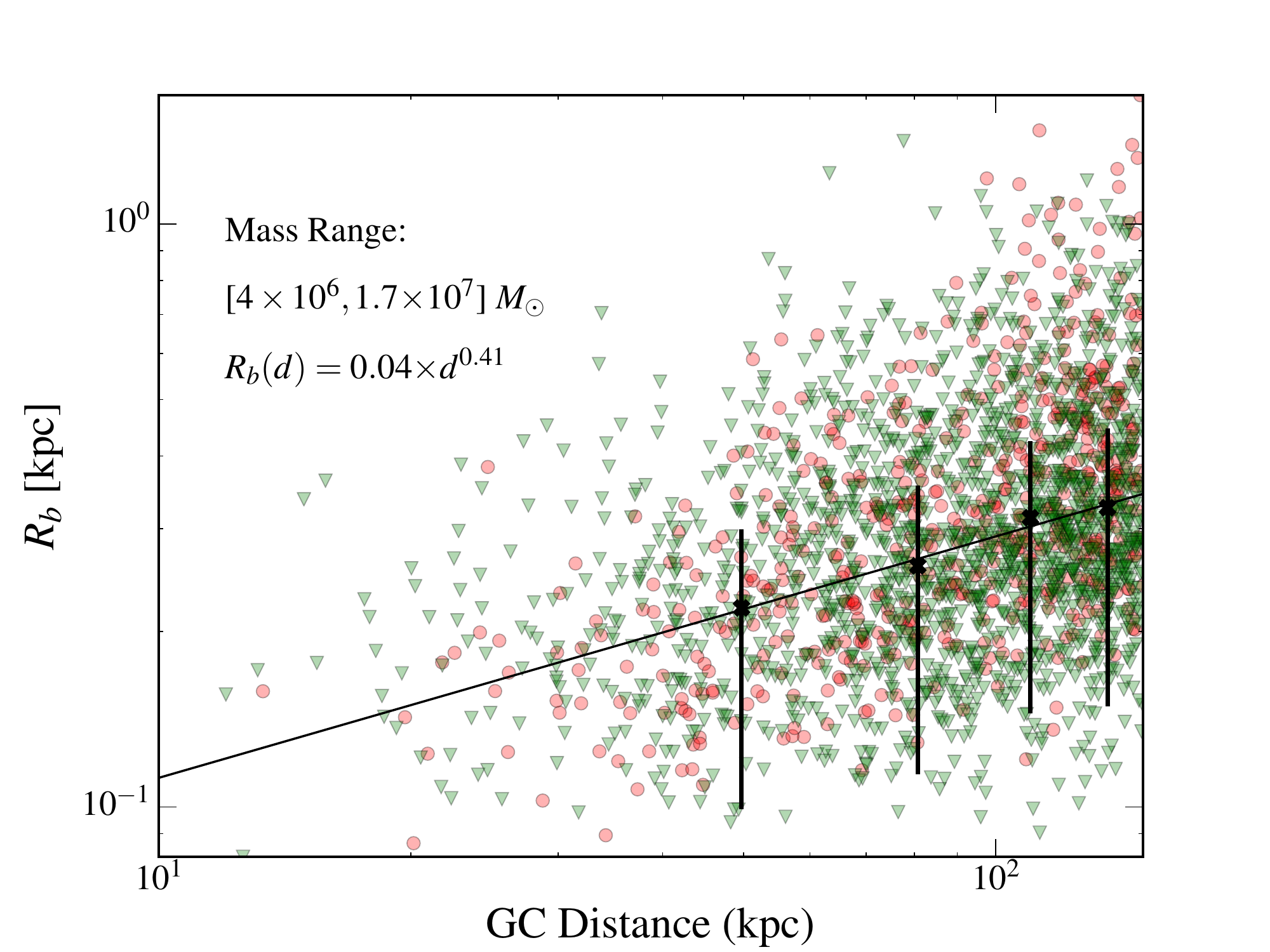}
\includegraphics[width=.49\textwidth, trim=0.4cm 0.0cm 0.4cm 0.4cm,clip=true]{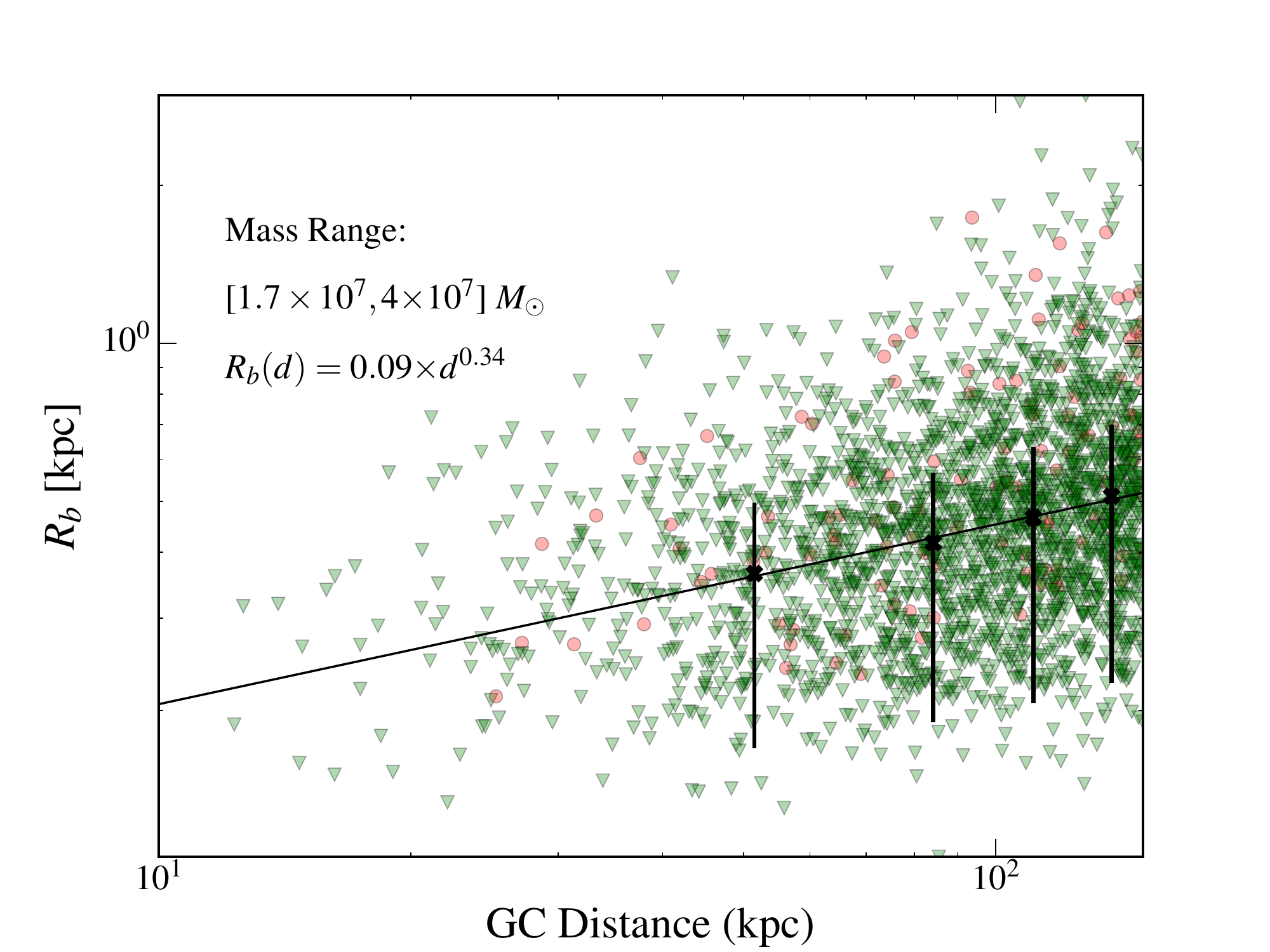}
\includegraphics[width=.49\textwidth, trim=0.4cm 0.0cm 0.4cm 0.4cm,clip=true]{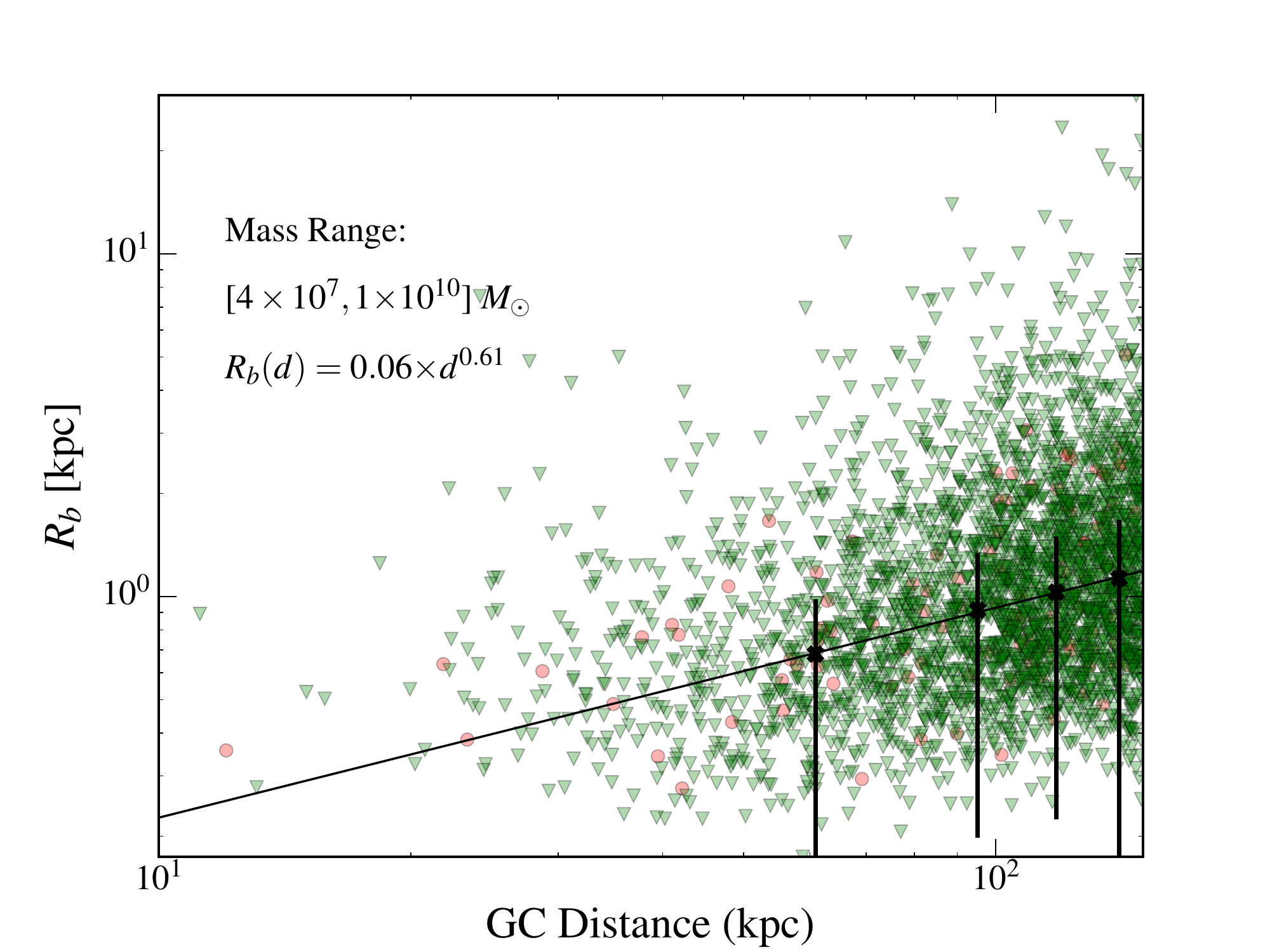}

\caption{\label{fig:scatter2} As in \Fig{fig:scatter1}, but for the parameter $R_b$, where $\rho(r) =\rho_0 \, r^{-\gamma} \exp(-r/R_b)$. As a result of tidal stripping, the average value of $R_b$ decreases with proximity to the center of the host halo. In each bin, we display the median best fit value of $R_b$, \ie $\langle R_b \rangle$, and $68\%$ containment region for each bin (denoted with a black `x' and vertical black lines respectively). The power law fit used to extrapolate the median $R_b$ values is shown in each panel, and is denoted $R_b(d)$. }
\end{figure*}

Our goal in this work is to identify the properties and distributions of the local subhalo population. Unfortunately, there are simply not enough subhalos in the inner tens of kiloparsecs to meaningfully extract properties exclusively from this sample. We approached this problem by identifying trends in the behavior of $\gamma$ and $R_b$ as functions of the total subhalo mass and the distance to the center of the host halo. This was accomplished by dividing subhalos into four mass bins, and then dividing each mass bin into four bins that differentiate halos by their distance to the Galactic Center. Bin sizes were chosen in such a way that each bin contains an approximately equal number of subhalos. Scatter plots of these best-fit values are shown in Figs.~\ref{fig:scatter1} and~\ref{fig:scatter2}.

As the result of tidal stripping, one can see in Fig.~\ref{fig:scatter2} that the average value of $R_b$ decreases with proximity to the center of the host halo. This result is consistent, for example, with the recent findings of Ref.~\cite{Moline:2016pbm} (see also e.g.~\cite{Hayashi:2002qv,Penarrubia:2010jk}). Perhaps less anticipated is that the average inner slope, $\gamma$, is also found to be lower for those subhalos located near the Galactic Center.

\begin{figure*}
\center
\vspace{1.0cm}
\includegraphics[width=0.495\textwidth]{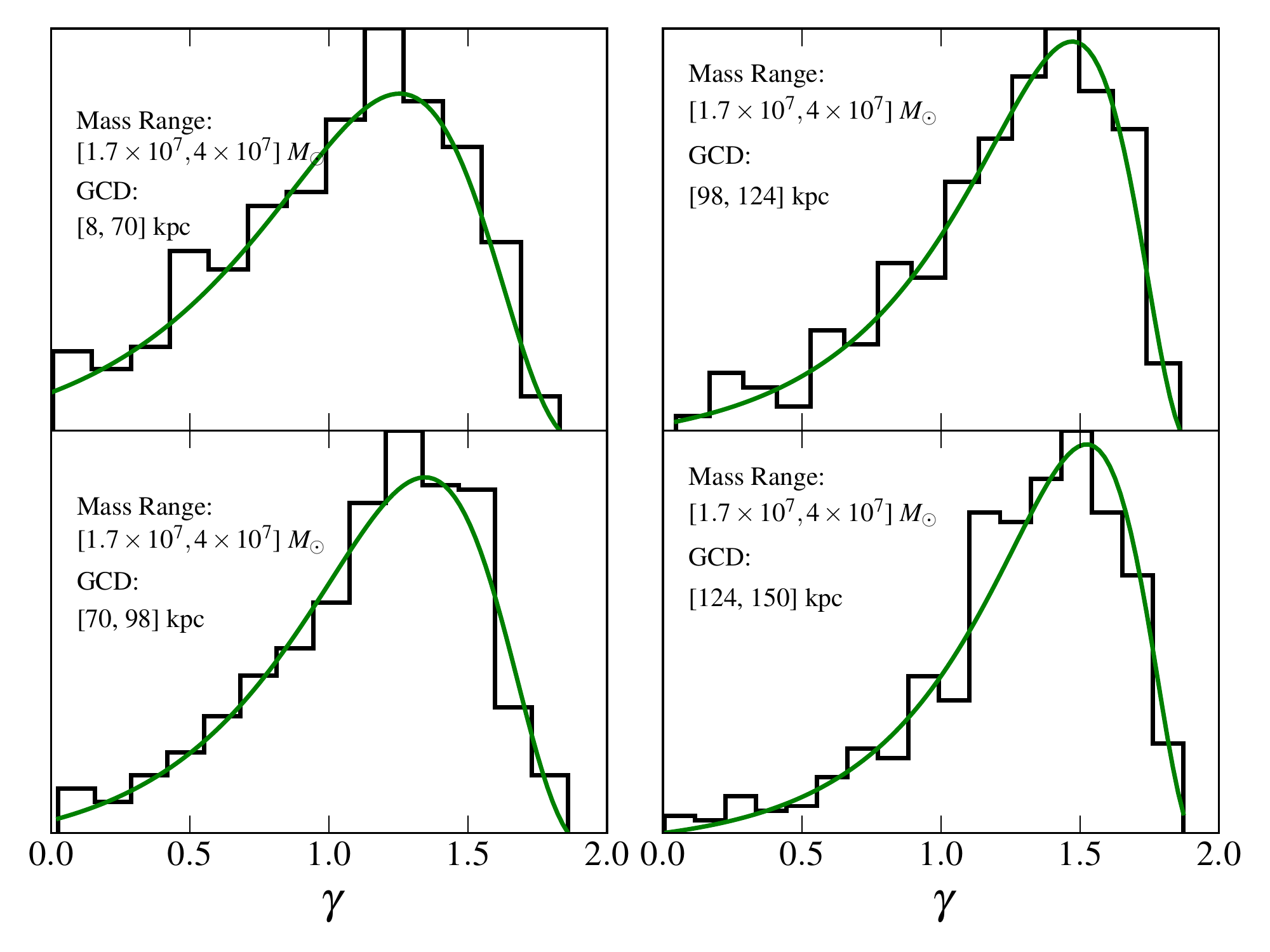}
\includegraphics[width=0.495\textwidth]{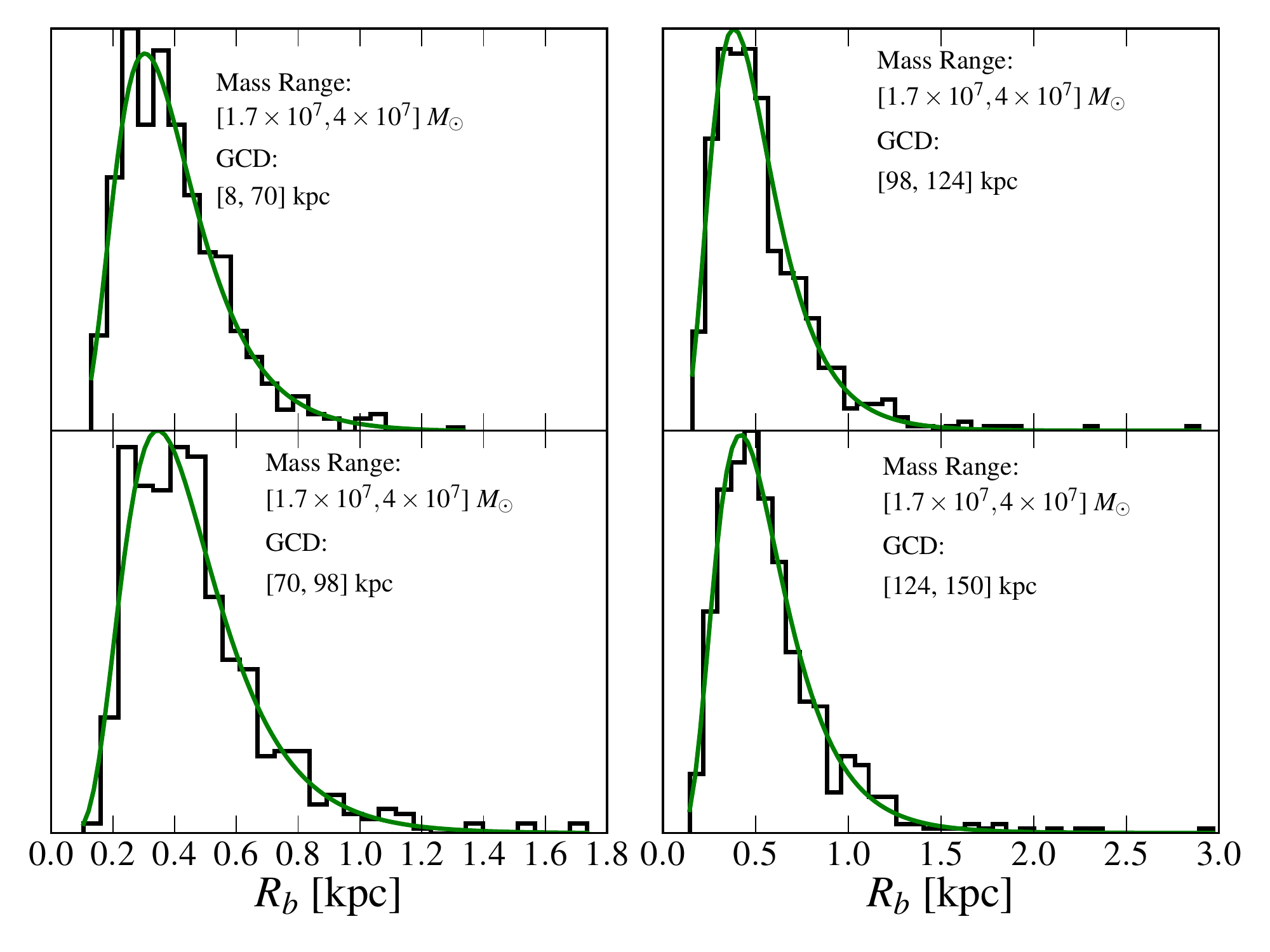}
\caption{\label{fig:gamma_rb_distribution}
The distribution of the inner slope (left) and exponential scale radius (right) for subhalos in the ELVIS and Via Lactea II simulations, for subhalos with masses in the range of $(2-5) \times 10^7 M_\odot$ and at various distances from the Galactic Center (GCD). The green line in each frame depicts the best-fit generalized normal (left) or lognormal (right) distribution.}
\end{figure*}

\begin{figure*}
\center
\includegraphics[width=0.495\textwidth, trim=1.0cm 0.0cm 0.8cm 0.4cm,clip=true]{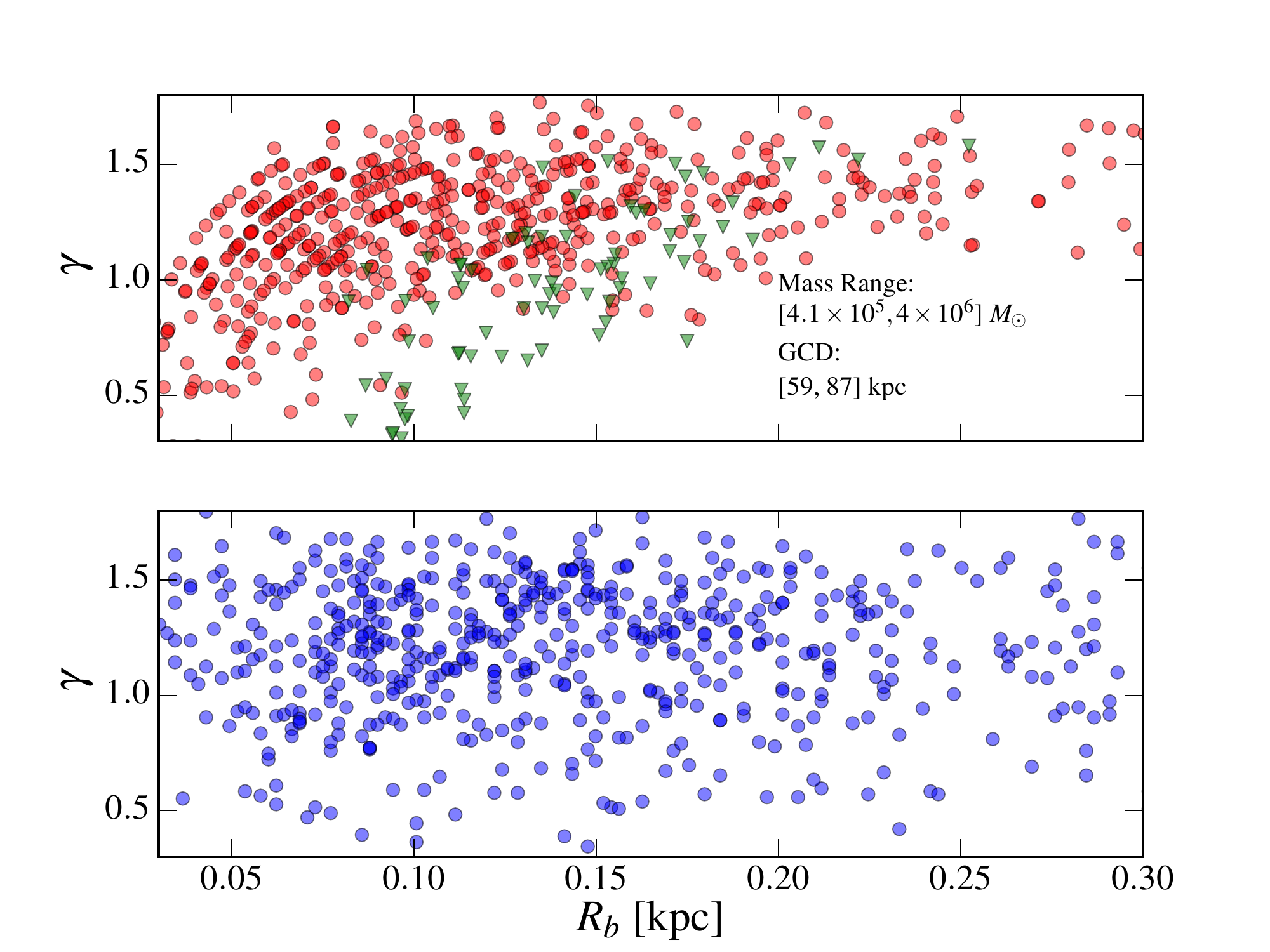}
\includegraphics[width=0.495\textwidth, trim=1.0cm 0.0cm 0.8cm 0.4cm,clip=true]{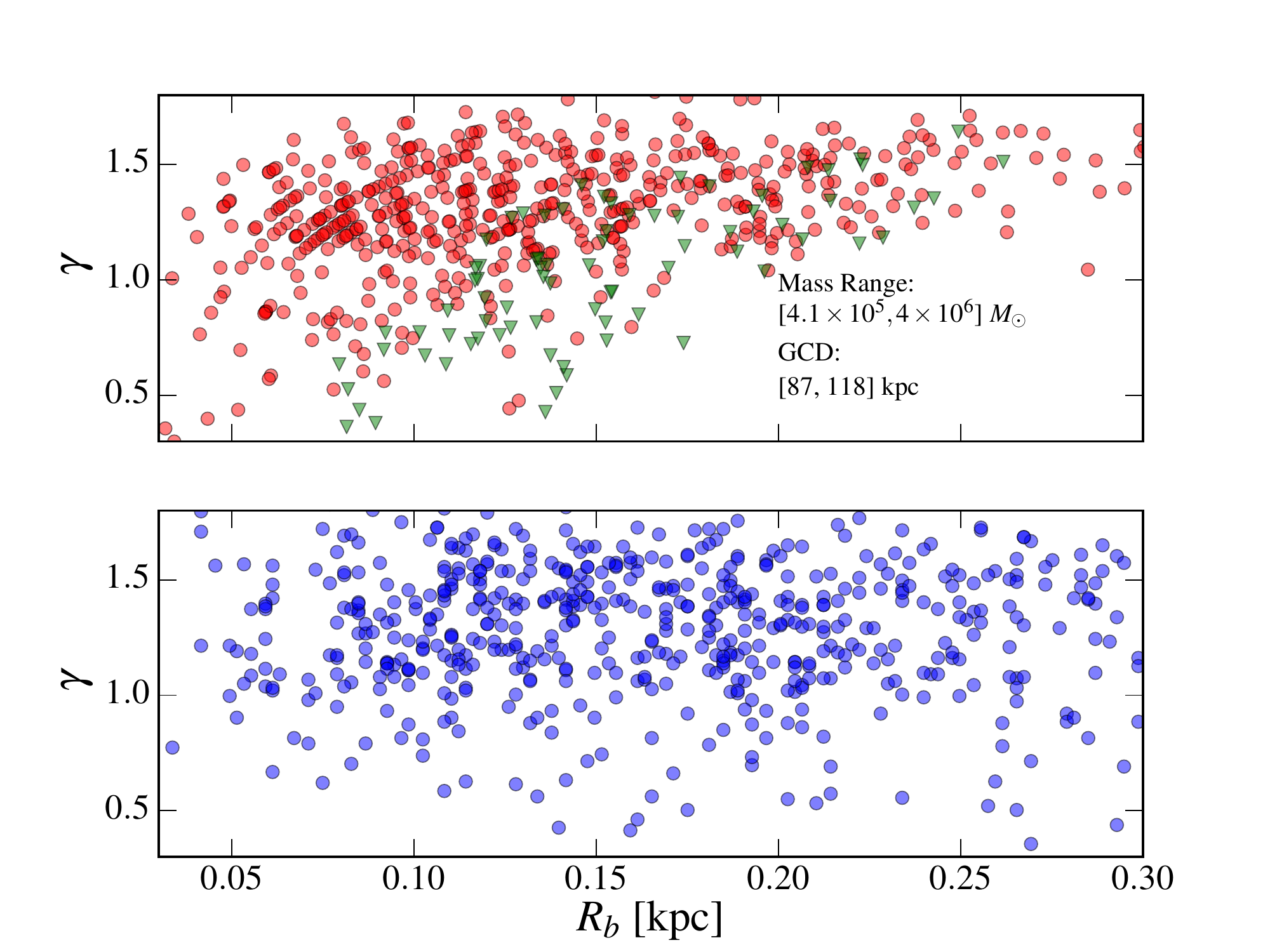}
\caption{\label{fig:correlations}
Comparison of best-fit parameters (top) $\gamma$ and $R_b$ to VL-II (red circles) and ELVIS subhalos (green triangles) and `fake' subhalos (bottom, blue) derived using random draws from \Eq{probrb} and \Eq{probgamma}. Analysis is shown for subhalo masses between $4.1 \times 10^{5}$ and $4 \times 10^6 M_\odot$, and for GCD ranges $[59, 87]$ kpc (left) and $[87, 118]$ kpc (right). } 
\end{figure*}

\begin{figure*}
\center
\includegraphics[width=0.495\textwidth, trim=0.4cm 0.0cm 0.4cm 0.4cm,clip=true]{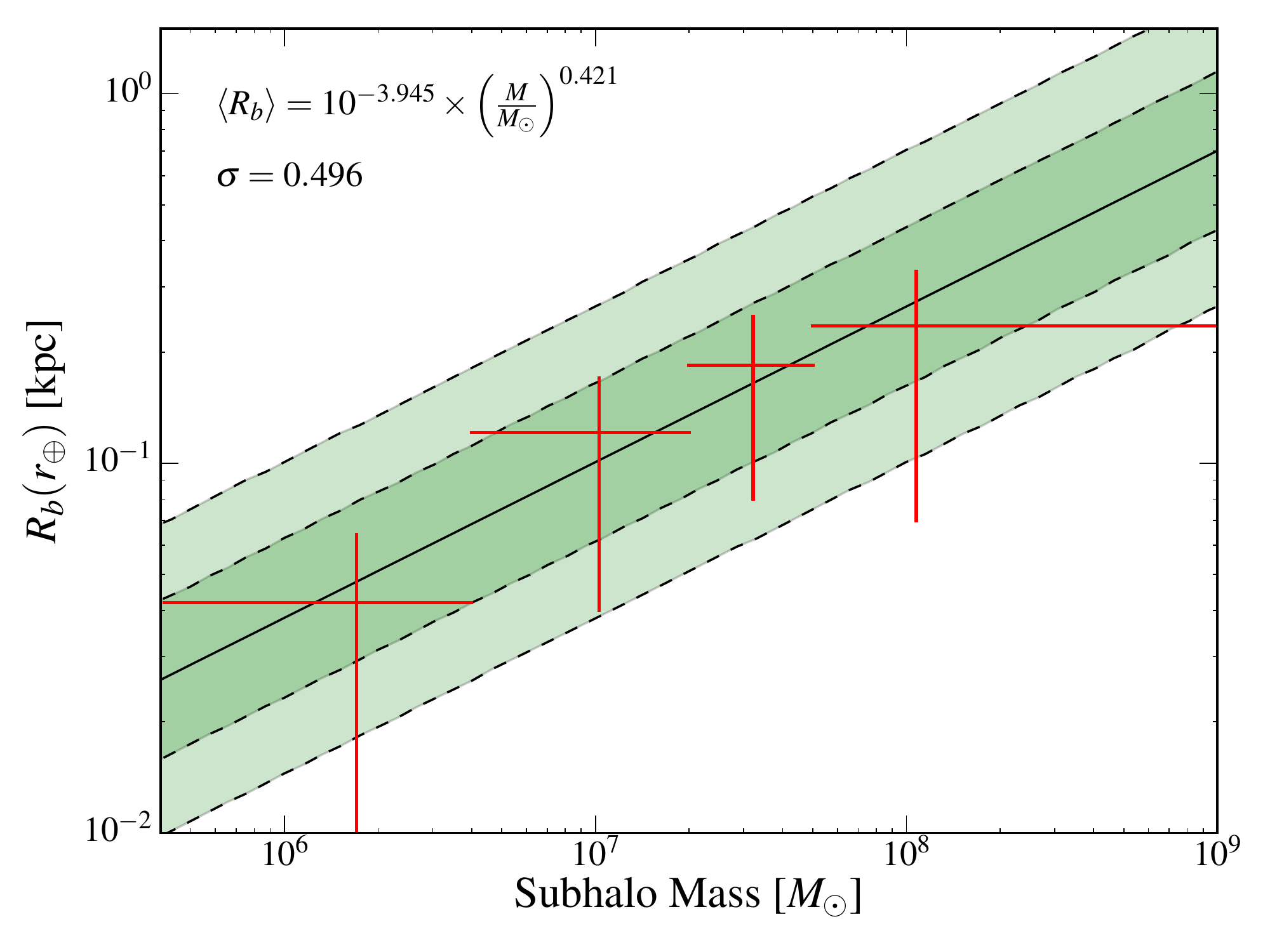}
\includegraphics[width=0.495\textwidth, trim=0.4cm 0.0cm 0.4cm 0.4cm,clip=true]{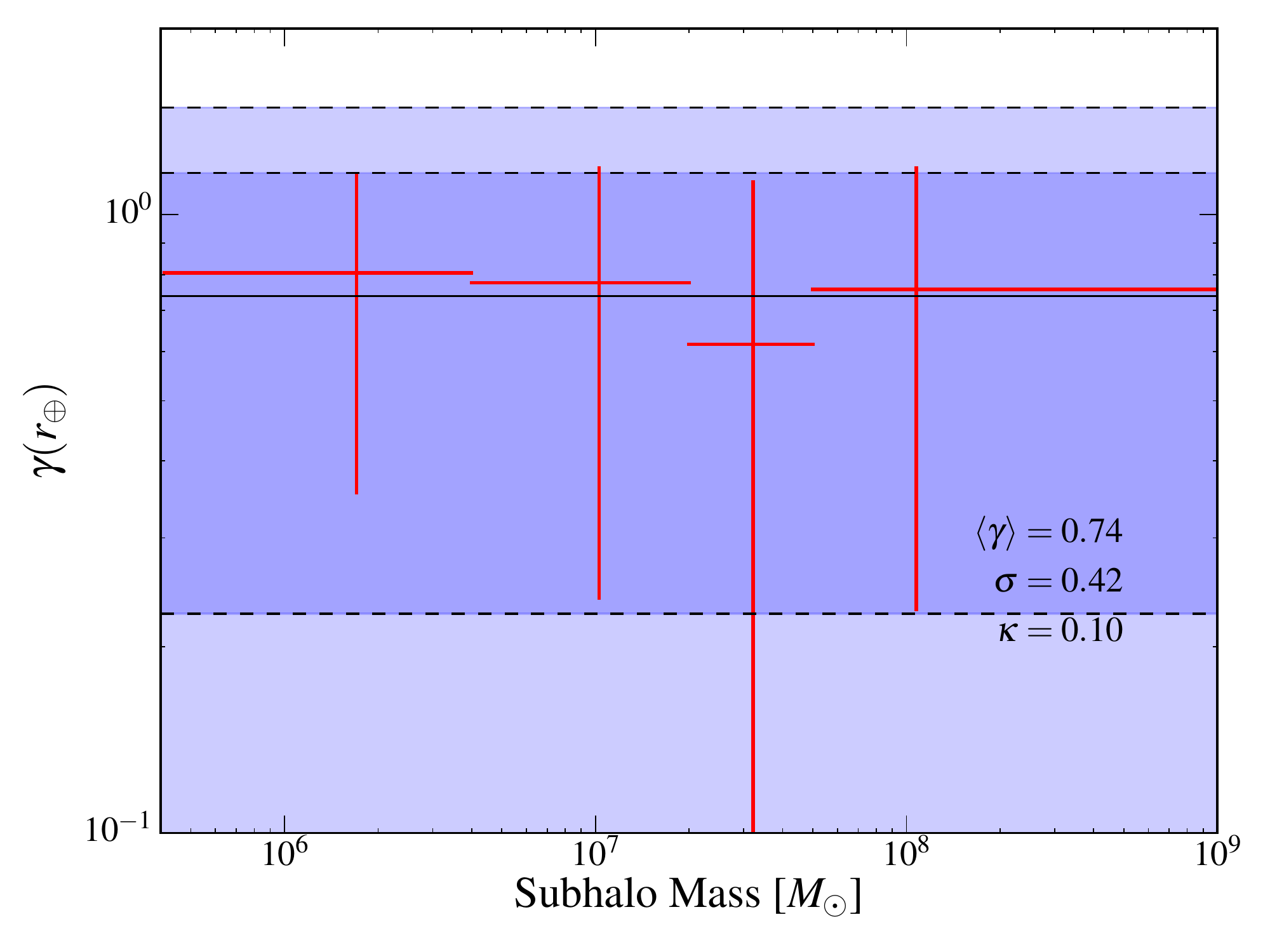}
\caption{\label{fig:fits}Left: The best-fit power law for the exponential scale parameter, $R_b$, as a function of subhalo mass, as determined from the ELVIS and Via Lactea II simulation data, for subhalos located $8.5$ kpc from the Galactic Center. Right: The median inner slope, $\gamma$, as determined from the ELVIS and Via Lactea II simulation data, for subhalos located $8.5$ kpc from the Galactic Center as a function of the subhalo mass. In each frame, the shaded regions depict the $68\%$ and $95\%$ containment contours.}
\end{figure*}

To parameterize the distribution of the values of $\gamma$  at Earth's location, we adopt a generalized normal distribution:
\begin{eqnarray}
\frac{dP}{d\gamma} &=& \frac{1}{\sqrt{2\pi}}\,\frac{1}{\sigma-\kappa(\gamma-\langle  \gamma \rangle)} \exp\bigg(-\frac{\ln^2( 1 - \kappa (\gamma - \langle  \gamma \rangle) / \sigma)}{2 \kappa^2} \bigg), \label{probgamma}
\end{eqnarray}
where $\langle \gamma \rangle$ is the median value of $\gamma$, and $\sigma$ and $\kappa$ are parameters which jointly characterize the width and skew of the distribution \footnote{For clarification, the width and skew are \emph{not} characterized by $\sigma$ and $\kappa$, respectively, but rather are more complicated functions of both of these parameters.}. Note that this distribution is defined on the domain $\gamma < \sigma/\kappa +\langle \gamma \rangle$. For each bin in subhalo mass and Galactic Center distance, we find the values of $\langle \gamma \rangle$, $\sigma$ and $\kappa$ which provide the best fit to the simulated dataset. Examples of the fitted distributions are shown for in the left panel of \Fig{fig:gamma_rb_distribution} for subhalo masses in the range of $(2-5) \times 10^7 M_\odot$ and various ranges of distance to the Galactic Center. For a fixed mass range, we fit a power law to the median value of gamma as a function of Galactic Center distance (the resultant fit equations for $\langle \gamma \rangle$ are given in each panel of \Fig{fig:scatter1}), as well as to the $\gamma$ values demarcating the edges of the $68\%$ containment region $\gamma_{68}^{+,-}$ --- defined as the values of $\gamma$ satisfying $\int_{\langle \gamma \rangle}^{\gamma_{68}^{+}}dP / d\gamma = 0.34$ and $\int_{\gamma_{68}^{-}}^{\langle \gamma \rangle}dP / d\gamma = 0.34$. The median \emph{local} value of $\gamma$ for a given range of subhalo masses is then determined using this power law fit. The equations defining the median power law fits for each mass range are provided in \Fig{fig:scatter1}. For each mass bin, the values of $\sigma$ and $\kappa$ characterizing the distribution in $\gamma$ are determined by requiring $\gamma_{68}^{+,-}$ equal the values determined by their respective power law extrapolations. Note that we take this approach, rather than attempting to extrapolate $\sigma$ and $\kappa$ directly, because there does not appear to be an obvious trend in either of these variables (this is a consequence of the fact that these variables do not independently correspond to physical features of the distribution). This procedure results in local $\gamma$ distributions for each of the $4$ distinct mass intervals. The median local value of gamma for each mass range, as well as the extrapolated $\gamma_{68}^{+,-}$ values, are shown in the right panel of \Fig{fig:fits}. Since this figure does not reveal any clear trend in the median value or in distribution of $\gamma$ as a function of subhalo mass, we parameterize the local $\gamma$ distribution as being independent of the subhalo mass, with values of $\langle \gamma \rangle$, $\sigma$, and $\kappa$ set to be the median of the local fits shown in \Fig{fig:fits} (the parameters of which are provided in \Fig{fig:fits}).

Similarly, to characterize the distribution of $R_b$ we adopt a log-normal distribution:
\begin{eqnarray}
\frac{dP}{dR_b} &=& \frac{1}{\sigma \sqrt{2\pi}} \, \frac{1}{R_b}\exp\bigg(-\frac{(\ln R_b - \ln \langle R_b \rangle)^2}{2 \sigma^2} \bigg),
\label{probrb}
\end{eqnarray}
where $\langle R_b \rangle$ is the median value of $R_b$, and $\sigma$ is the width of the distribution. Once again, we find the values of $\langle R_b \rangle$ and $\sigma$ which provide the best fit to the simulated dataset in each bin in subhalo mass and Galactic Center distance. We find that the value of $\sigma$ is not dependent on the subhalo mass or Galactic Center distance, and thus we average the preferred value across all bins. Examples of the fitted distributions are shown for in the right panel of \Fig{fig:gamma_rb_distribution} for subhalo masses in the range of $(2-5) \times 10^7 M_\odot$ and various ranges of distance to the Galactic Center. The local median value of $R_b$ for each mass bin is determined using the power law fits to the median $R_b$ value of each bin, shown in \Fig{fig:scatter2} (these equations are also provided in \Fig{fig:scatter2} for each mass interval). The local $\langle R_b \rangle$ fit and $68\%$ containment regions for each subhalo mass bin are shown in the left panel of \Fig{fig:fits}. We find that the mass dependence of the exponential scale parameter is well described by a power-law, with $\langle R_b(M) \rangle \propto M^{0.421}$. We thus use this power law, along with the averaged $\sigma$ value, to characterize the mass dependence of the local distribution of $R_b$ (the final parameters characterizing this distribution are provided in left panel of \Fig{fig:fits}). We have verified that our derived distributions in both $R_b$ and $\gamma$ are relatively insensitive to the choice of binning. 

To address possible correlations between $R_b$ and $\gamma$ that are not captured by our one dimensional parameterizations, we compare in the $R_b-\gamma$ plane the VL-II and ELVIS best-fit parameters to the best-fit parameters that would be derived from randomly drawing values of $R_b$ and $\gamma$ from \Eq{probrb} and \Eq{probgamma}, assuming subhalo characteristics (mass and GC distance) are identical to those of the VL-II and ELVIS subhalos. The result of this test is shown in \Fig{fig:correlations} for two different bins. The independent extrapolations appear to do a very reasonable job of capturing the subhalo properties. \Fig{fig:correlations} does suggest, however, that our distributions may slightly over-estimate the number of small-$\gamma$ large-$R_b$ subhalos and the number of large-$\gamma$ small-$R_b$ subhalos. It is difficult to the assess the overall impact of this mis-modeling given that these overestimations lead to opposite effects. In \Sec{sec:uncert} we will demonstrate the extent to which reducing halo-to-halo variations impacts the number of observable subhalos; from there, one many attempt to infer the effect that this mis-modeling may have on the derived limits.

\begin{figure*}
\center
\includegraphics[width=\textwidth]{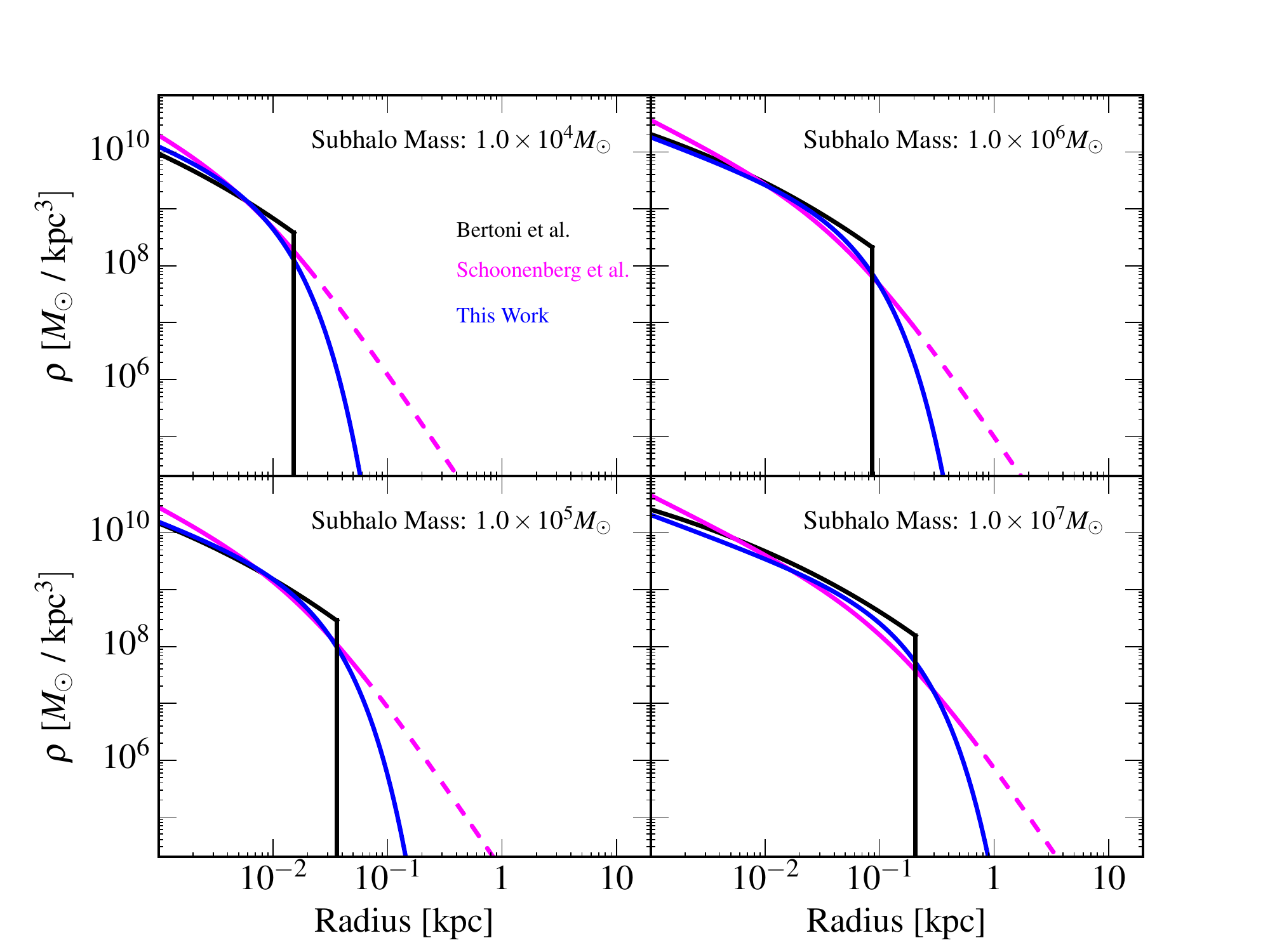}
\caption{\label{densitycompare}
A comparison of the median density profiles adopted in Ref.~\cite{Bertoni:2015mla} (black), Ref.~\cite{Schoonenberg:2016aml} (magenta), and as derived in this study (blue) and for subhalos of four different masses.}
\end{figure*}

If Fig.~\ref{densitycompare}, we compare the median subhalo density profiles adopted in Ref.~\cite{Bertoni:2015mla} (black), Ref.~\cite{Schoonenberg:2016aml} (magenta), and as derived in this study (blue), for subhalos ranging in mass from $10^4 M_\odot$ to $10^7 M_\odot$. As the volume integral of the NFW profile adopted in Ref.~\cite{Schoonenberg:2016aml} yields a subhalo mass that exceeds that reported by VL-II, we take the outer regions of the dark matter distribution to be unspecified in this case. To convey this, we plot this profile as a solid line only within the radius that contains the mass reported by VL-II, and as a dashed line beyond this point.\footnote{We note that Ref.~\cite{Schoonenberg:2016aml} does not explicitly state how they reconstruct density profiles at fixed subhalo mass from the extracted VL-II simulation results. The curves shown here are the result of a cubic spline interpolation function fit to $\ln R_{v,{\rm max}}(\ln M)$ and $\ln v_{c,{\rm max}}(\ln M)$ in the mass range where VL-II can resolve subhalos, and that we extrapolate to lower masses using a power law fit. The results obtained in this fashion appear to be quite similar to those presented in Ref.~\cite{Schoonenberg:2016aml}.} 


\begin{figure*}
\center
\includegraphics[width=0.495\textwidth]{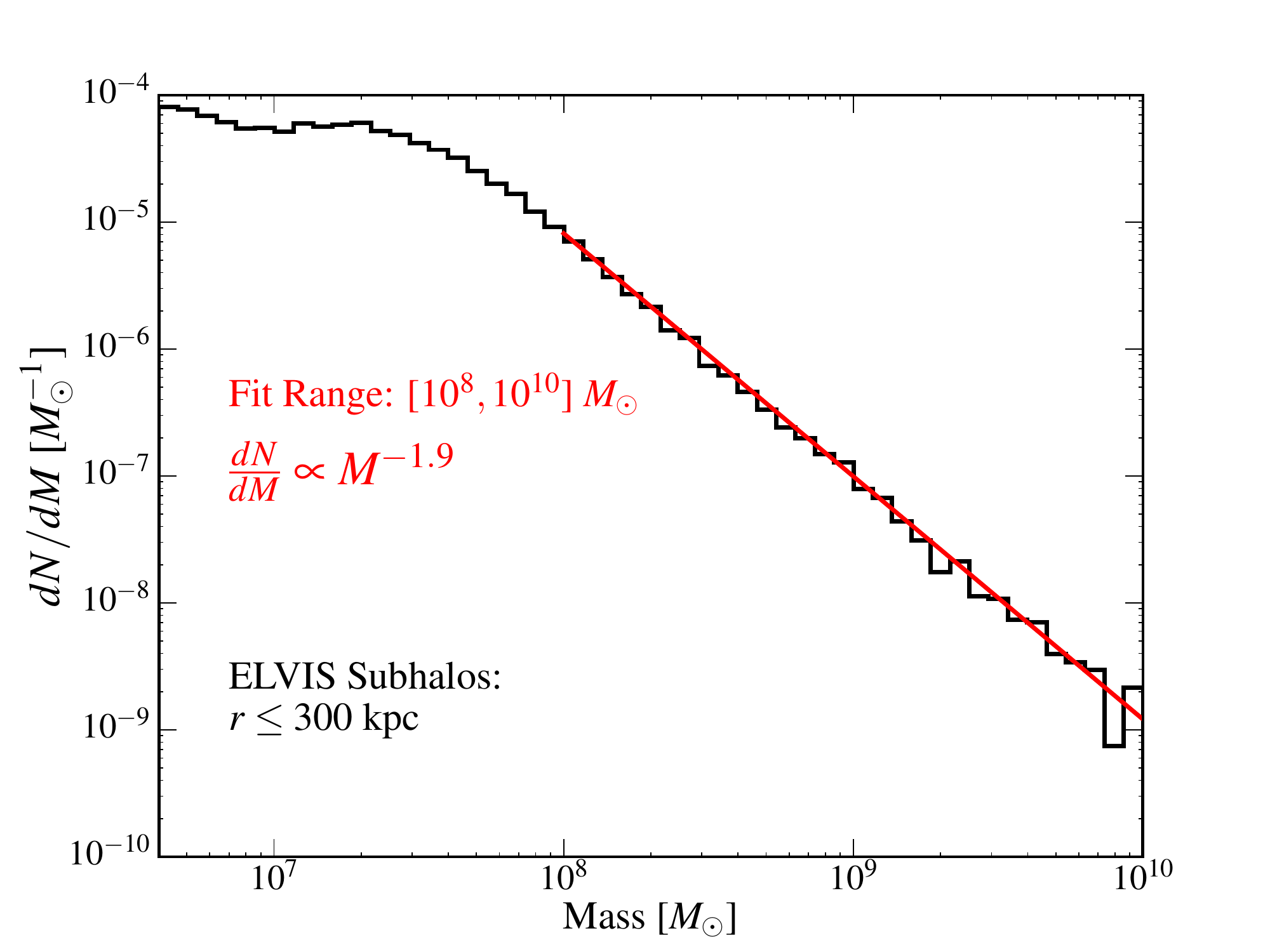}
\includegraphics[width=0.495\textwidth]{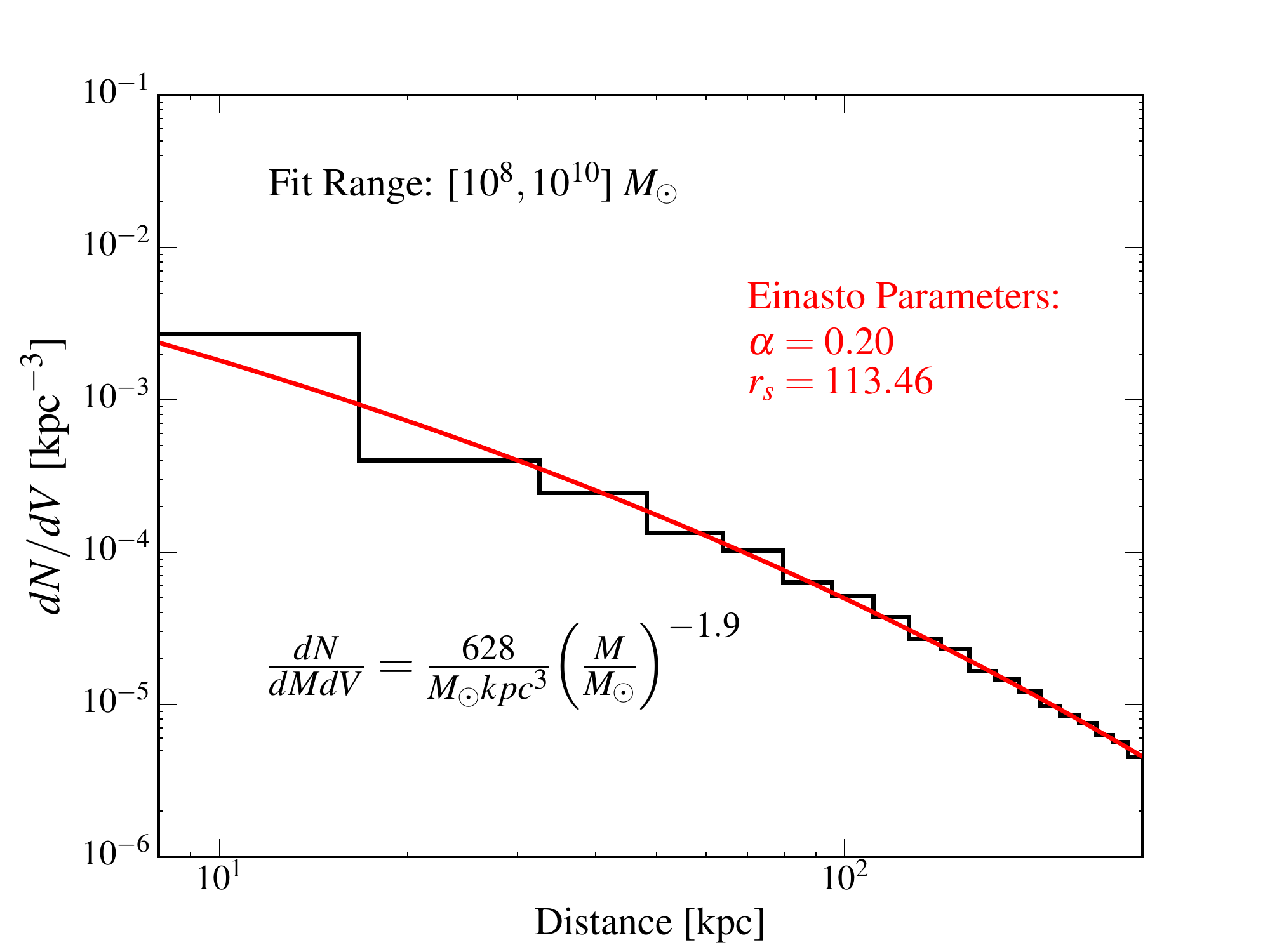}
\caption{\label{fig:subnumdense}
Left: The number of subhalos per unit mass within 300 kpc of the host halo's center as found in the ELVIS suite of simulations. The red line represents the best-fit power-law for subhalos with mass between $10^8$ and $10^{10} \, M_{\odot}$. Right: The number of subhalos per unit volume in the ELVIS simulations for subhalos with masses between $10^8$ and $10^{10} M_\odot$. The red line depicts the best-fit Einasto profile for this subhalo distribution. At a distance of 8.5 kpc from the center of the host halo, this corresponds to a local population described by $dN/dMdV = 628 \, {\rm kpc}^{-3} M_{\odot}^{-1} \times (M/M_{\odot})^{-1.9}$.}
\end{figure*}

In order to predict the number of subhalos that could be observed by Fermi-LAT, one needs not only the shapes of the subhalo density profiles, but also the local number density of subhalos of a given mass. 
%
%
In the left panel of Fig.~\ref{fig:subnumdense}, we plot the number of subhalos in the ELVIS simulation per unit subhalo mass as a function of subhalo mass. For masses above $10^8 \, M_{\odot}$, we find that this distribution is well fit by $dN/dM \propto M^{-1.9}$, consistent with previous literature~\cite{Springel:2008cc, Diemand:2008in, Garrison-Kimmel:2013eoa, Despali:2016meh}. Although the distribution appears to depart from this power-law at lower masses, we attribute this to the finite resolution of the ELVIS simulation.


In the right panel of Fig.~\ref{fig:subnumdense}, we plot the number density of subhalos as a function of the distance to the host center. Consistent with Refs.~\cite{Springel:2008cc, Despali:2016meh}, we find that this distribution is well-characterized by an Einasto profile. We use the fitted Einasto profile rather than the histogram itself to extract the local number density to avoid sensitivity to the choice of binning. This allows us to derive the following distribution for the local subhalo population:
\begin{equation}\label{eq:numdens}
\frac{dN}{dM dV} = \frac{628}{M_\odot \, {\rm kpc}^3}\left(\frac{M}{M_\odot}\right)^{-1.9} \, .
\end{equation}
In principle, a complete subhalo analysis would use the full radial dependence of the subhalo number density. However, we find that for the cross sections considered, effectively all observable subhalos reside very near to Earth where the subhalo number density is approximately constant. Note that had we considered more massive subhalos (\eg dwarf galaxy sized objects), this would no longer have been the case and the radial dependence of the number density would be important. We thus approximate the number density as a location independent function using \Eq{eq:numdens}. We caution the reader that while this approximation is not thought to introduce significant error, it is possible that it results in a slight overestimation of the number of observable subhalos. 

In the following section, we will use this subhalo distribution, along with the afore described distribution of subhalo density profiles, to calculate the gamma-ray luminosity function of local subhalos, and in turn the number of such subhalos that are predicted to be detectable to Fermi and other gamma-ray telescopes. 


\section{Detecting Dark Matter Subhalos With Gamma-Ray Telescopes \label{sec:detection}}

\subsection{Gamma-Rays from Dark Matter Subhalos}

A given subhalo will generate a gamma-ray flux that is given by:
\begin{equation}\label{eq:gamma_ray_flux}
\Phi_{\gamma} = \frac{\langle \sigma v \rangle N_{\gamma}}{8 \pi m^2_\chi D^2} \int \rho^2(r) \, dV,
\end{equation}
where $\left<\sigma v \right>$ is the dark matter's thermally averaged self-annihilation cross section, $N_\gamma$ is the number of gamma rays produced per annihilation, $m_\chi$ is the dark matter mass, $D$ is the distance to the center of the subhalo, and $\rho(r)$ is the density profile of the subhalo. For a given dark matter mass and annihilation channel, we calculate $N_{\gamma}$ using Pythia 8~\cite{pythia8}.


From the elements described in the previous section, we can calculate the number of subhalos that yield a gamma-ray flux above a given flux threshold, $\Phi_{\rm Thresh}$: 
\begin{eqnarray}\label{eq:totnum}
N_\text{obs}  =  \Omega \int \int \int \int D^2 \, \frac{dN}{dMdV} \,  \frac{dP}{d\gamma} \,\frac{dP}{dR_b} \, \Theta[\Phi_{\gamma}(M, D, R_b, \gamma)-\Phi_{\rm Thresh}]\, dM \, dD \,  dR_b \, d\gamma,  \nonumber \\
\end{eqnarray}
where $dN/dM dV$ is the local subhalo number density per unit mass (Eq.~\ref{eq:numdens}) and $dP/d\gamma$ and $dP/dR_b$  are the generalized normal and lognormal distributions for the parameters $\gamma$ and $R_b$, respectively (Eqns.~\ref{probgamma} and~\ref{probrb}). The quantity $\Omega$ is the solid angle observed, which in the case of $|b|>20^{\circ}$ corresponds to $4\pi (1-\sin 20^{\circ})$. We choose to limit the parameter $\gamma$ to the range of 0 to 1.45, re-normalizing the distribution such that $\int ^{1.45}_0 (dP/d\gamma) \, d\gamma =1$. This is done for two reasons. First, subhalos with $\gamma < 0$ are unlikely to be physical, as there are no mechanisms at play in these dark matter only simulations which should cause the density profile to increase as a function of radius. Note that this truncation occurs only at the far tail of the distribution, resulting in a small effect. The far more important effect is truncating $\gamma > 1.45$. The density integral in \Eq{eq:gamma_ray_flux} is divergent for $\gamma \geq 1.5$. This is not to say such halos cannot be physical, only that the density distribution must develop a core at some inner radii. In order to avoid having to specify the specific nature of such a core, we remove this part of the distribution. This truncation is conservative as subhalos with larger $\gamma$ produce a noticeably larger flux, and are thus more observable. In order to compare our calculations to the list of subhalo candidates in the 3FGL gamma-ray source catalog as presented in Ref.~\cite{Bertoni:2015mla}, we adopt a value of $\Phi_{\rm Thresh}=7\times 10^{-10}$ cm$^{-2}$ s$^{-1}$ and consider only photons with energies above 1 GeV. We restrict our attention to subhalos with masses below $10^7\, M_{\odot}$ to avoid the inclusion of any dwarf galaxies and treat the minimum subhalo mass as a free parameter.

We note that because our analysis focuses on local subhalos and explicitly approximates the subhalo number density as independent of GC distance, $\log_{10}(N_\text{obs}) \propto \frac{3}{2}\log_{10}(\left<\sigma v \right>)$. This scaling relation differs from various published results, particularly for analyses that include dwarf-sized objects (see \eg Fig. 5 of Ref.~\cite{Schoonenberg:2016aml} and Fig. 9 of Ref.~\cite{Calore:2016ogv}). This is because dwarf-sized objects can be observed at much larger distances where the constant number density approximation may no longer be valid. 


\subsection{Placing Constraints on the Dark Matter Annihilation Cross Section}

Analyses of the unidentified sources in Fermi's 3FGL catalogue have identified 19 bright ($\Phi_{\gamma} >7\times 10^{-10}$ cm$^{-2}$ s$^{-1}$ ), high-latitude ($|b| > 20^\circ$) sources with no evidence of variability and which exhibit a spectral shape consistent with annihilating dark matter~\cite{Bertoni:2015mla, Bertoni:2016hoh}.\footnote{The 19 subhalo candidates are the same as those listed in Ref.~\cite{Bertoni:2015mla}, after removing the five sources that have more recently been associated with emission at other wavelengths~\cite{Bertoni:2016hoh}.}
 In this subsection, we will use the observed number and characteristics of these subhalo candidates to place upper limits on the dark matter annihilation cross section. 

Following the approach of Ref.~\cite{Bertoni:2015mla}, we calculate the $\chi^2$ associated with the fit of a given dark matter model to the spectrum of each subhalo candidate, and define the weighted number of sources (WNS) to be twice the sum of the $p$-values associated with the fit, \ie 
\begin{equation}
{\rm WNS} \equiv 2 \sum_{\rm i} p_i = 2 \sum_{\rm i} \int_{\chi^2_{{\rm obs}, i}}^{\infty} f_{k}(x) \, dx 
\end{equation}
(refer to the left panel of Fig.~10 in Ref.~\cite{Bertoni:2015mla} for result). Here, $p_i$ is the $p$-value associated with source $i$, $f_{k}(x)$ is the $\chi^2$ distribution function for $k$ degrees of freedom, and $\chi^2_{{\rm obs},i}$ is the observed chi-square value of source $i$. We then apply Poisson statistics to the WNS to place a 95\% upper limit on the annihilation cross section, for a given value of the dark matter mass and annihilation channel. 

\begin{figure*}
\center
\includegraphics[width=0.7\textwidth]{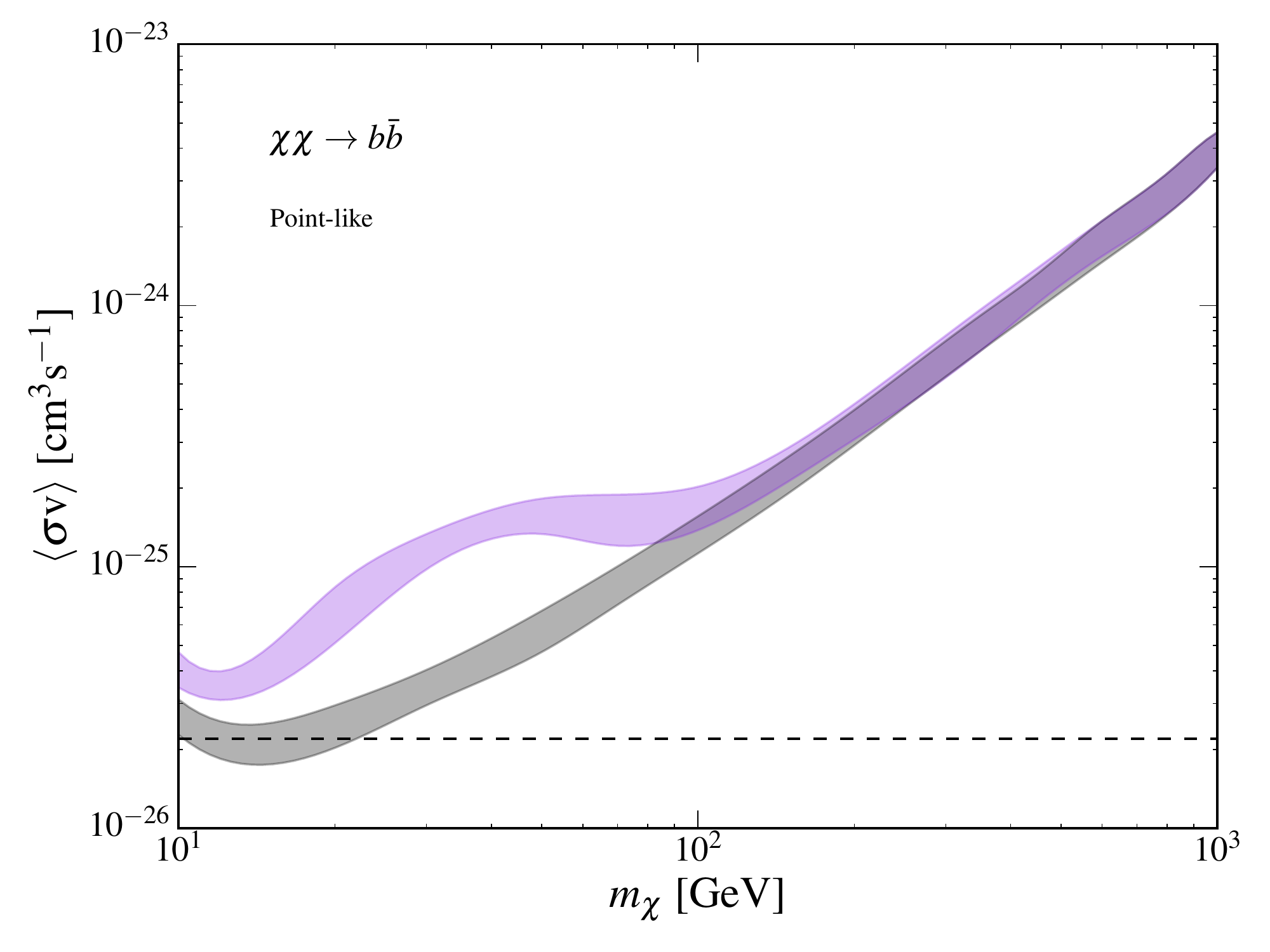}
\caption{\label{fig:limbb}
The $95\%$ confidence level upper limit on the cross section for dark matter annihilation to $b\bar{b}$ as derived from the unassociated gamma-ray source population presented in \Ref{Bertoni:2015mla} (purple). Also shown are the limits that would have been derived if no subhalo candidates been observed (grey). }
\end{figure*}

\begin{figure*}
\center
\includegraphics[width=\textwidth]{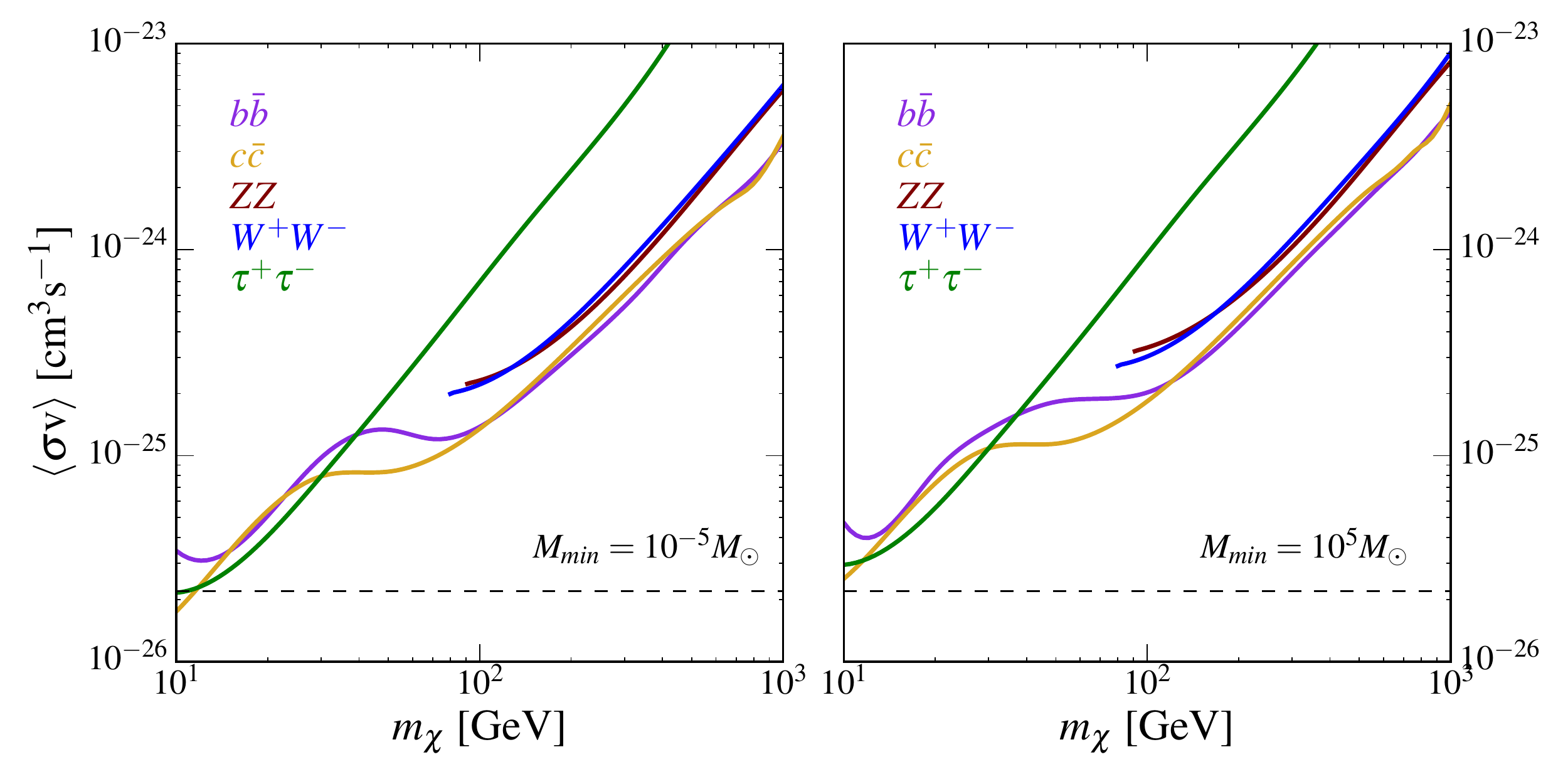}
\caption{\label{fig:limall}
The $95\%$ confidence level upper limit on the dark matter annihilation cross section for various annihilation channels, adopting a minimum subhalo mass of $10^{-5} M_\odot$ (left) or $10^5 M_\odot$ (right).}
\end{figure*}

In Fig.~\ref{fig:limbb}, we plot the upper limit derived for dark matter annihilating to $b\bar{b}$ (purple). The upper (lower) boundary of this band represents the limit obtained assuming a minimum subhalo mass of $10^5 \,  M_{\odot}$ ($10^{-5}M_\odot$). We also show in this figure the limits that would have been obtained if no subhalo candidate sources had been detected (zero weighted sources). As a benchmark, we plot as a dashed horizontal line the cross section associated with dark matter in the form of a simple thermal relic. In Fig.~\ref{fig:limall}, we show the $95\%$ upper limits for dark matter annihilating to various final states ($b\bar{b}, \, c\bar{c}, \, \tau^+\tau^-, \, ZZ$ or $W^+W^-$), adopting a minimum mass of either $10^{-5}M_\odot$ (left) or $10^{5}M_\odot$ (right).

\begin{figure*}
\center
\includegraphics[width=0.7\textwidth]{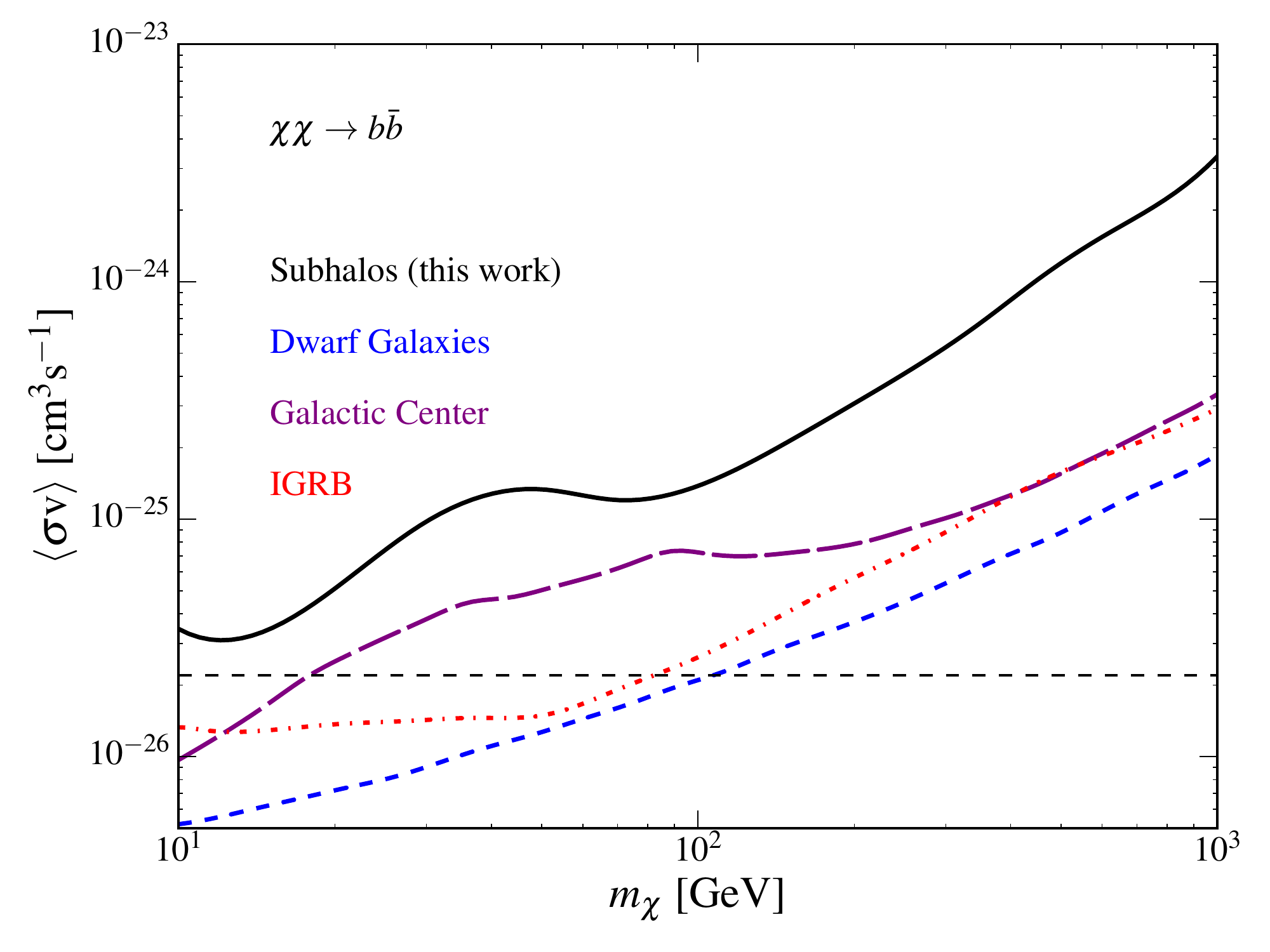}
\caption{\label{fig:external}
A comparison of the $95\%$ confidence level upper limits on the dark matter annihilation cross section placed from gamma-ray searches for subhalos (solid black) and gamma-ray observations of dwarf galaxies (short-dashed blue)~\cite{Drlica-Wagner:2015xua}, the Galactic Center (long-dashed purple)~\cite{Hooper:2012sr} and the isotropic gamma-ray background (dot-dashed red)~\cite{Ackermann:2015tah}. Here we have adopted a minimum subhalo mass of $10^{-5} M_\odot$ and consider the representative case of annihilations to $b\bar{b}$.}
\end{figure*}

In Fig.~\ref{fig:external}, we compare the limits on the annihilation cross section derived in this study to those previously obtained from Fermi's observations of dwarf galaxies (short-dashed blue)~\cite{Drlica-Wagner:2015xua}, the Galactic Center (long-dashed magneta)~\cite{Hooper:2012sr} and the isotropic gamma-ray background (dot-dashed green)~\cite{Ackermann:2015tah}. Although the limits from subhalo searches are somewhat weaker than those derived from these other observational targets, they are reasonably competitive and highly complementary.

\subsection{Prospects for Detecting Spatial Extension}

Thus far our discussion has been restricted to the detection of dark matter subhalos as point-like gamma-ray sources. Of those subhalos detectable by Fermi, however, the most massive and nearby may be discernibly spatially extended, potentially enabling one to distinguish a dark matter subhalo from a pulsar, blazar, or other gamma-ray point source.  The unambiguous observation of a spatially extended gamma-ray source with no corresponding emission at other wavelengths would constitute a smoking gun for annihilating dark matter~\cite{Bertoni:2016hoh}. 

To quantify the degree of spatial extension of the gamma-ray emission from a dark matter subhalo, we introduce the parameter, $\sigma_{_{68}}$, defined as the angular radius which contains $68\%$ of the total photons from the source:
\begin{equation}\label{eq:sig68}
\frac{\int_0^{\sigma_{_{68}}}\, \rho(r)^2 \,dl}{\int_{0}^{\theta_\text{max}}\, \rho(r)^2 \, dl} = 0.68 ,
\end{equation}
where the integrals are performed over the line-of-sight, and $\theta_\text{max}$ is the angular radius encompassing the full extension of the subhalo. Given the point spread function of Fermi, a bright unassociated source can be potentially identified as spatially extended if $\sigma_{_{68}} \gsim \mathcal{O}(0.1^{\circ})$~\cite{Bertoni:2016hoh}. In the case of bright point-like gamma-ray sources, Fermi can typically place upper limits on the degree of spatial extension at approximately the same level.

\begin{figure*}
\center
\includegraphics[width=0.7\textwidth]{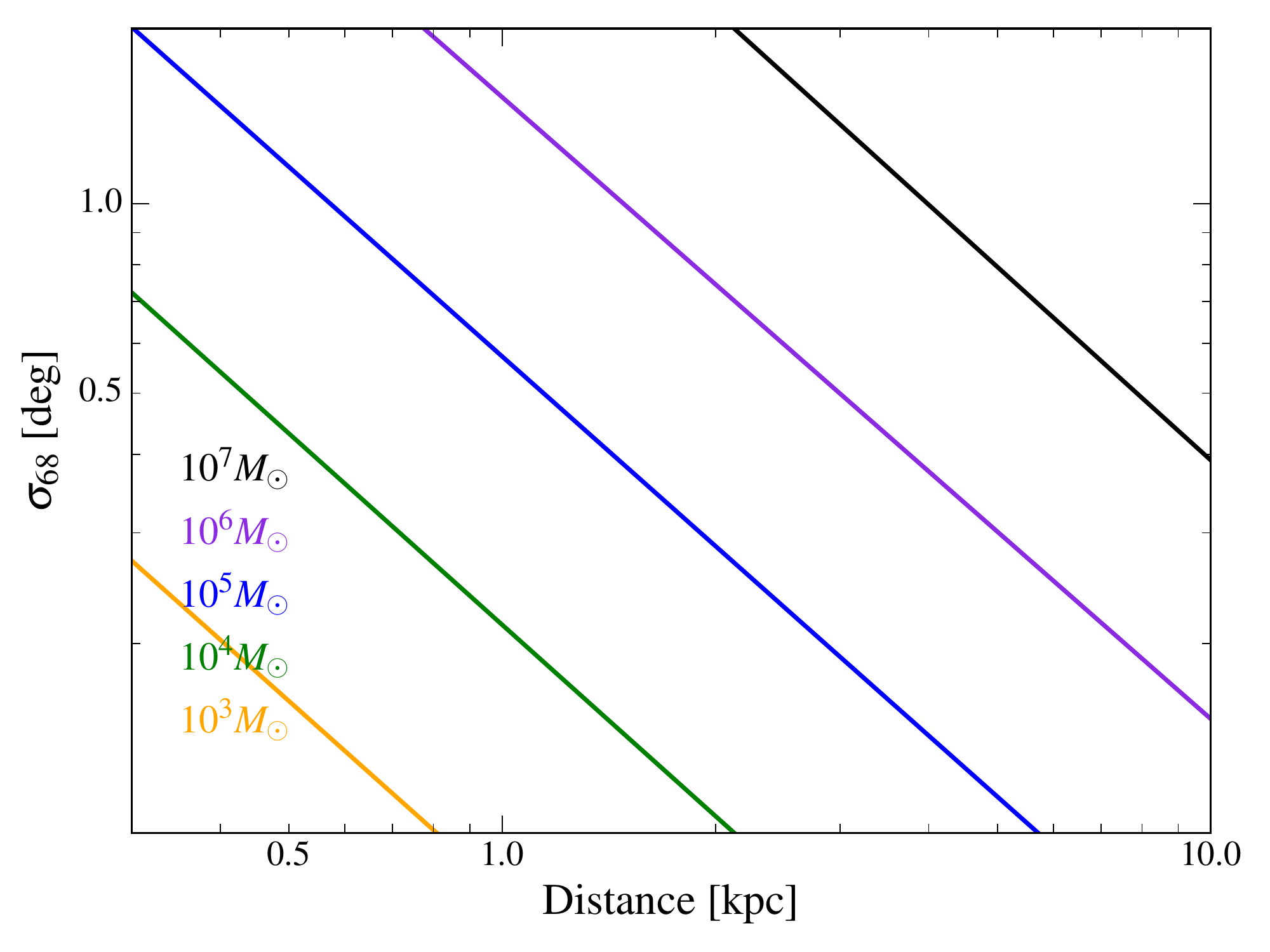} 
\caption{\label{fig:extension}
The 68\% containment radius, $\sigma_{_{68}}$, as a function of the distance to a given subhalo, for five values of the subhalo mass.}
\end{figure*}

\begin{figure*}
\center
\includegraphics[width=.49\textwidth]{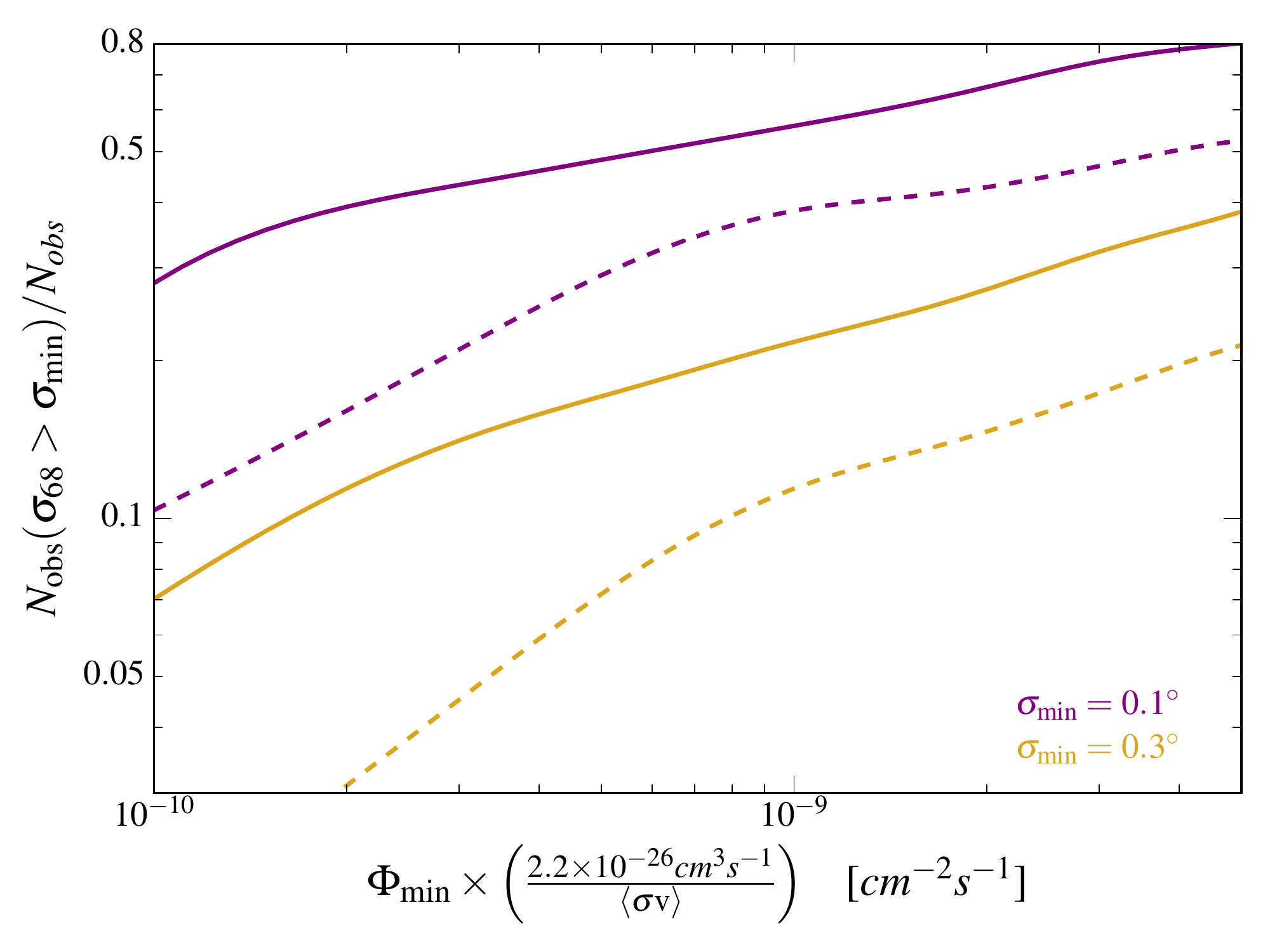}
\includegraphics[width=.49\textwidth]{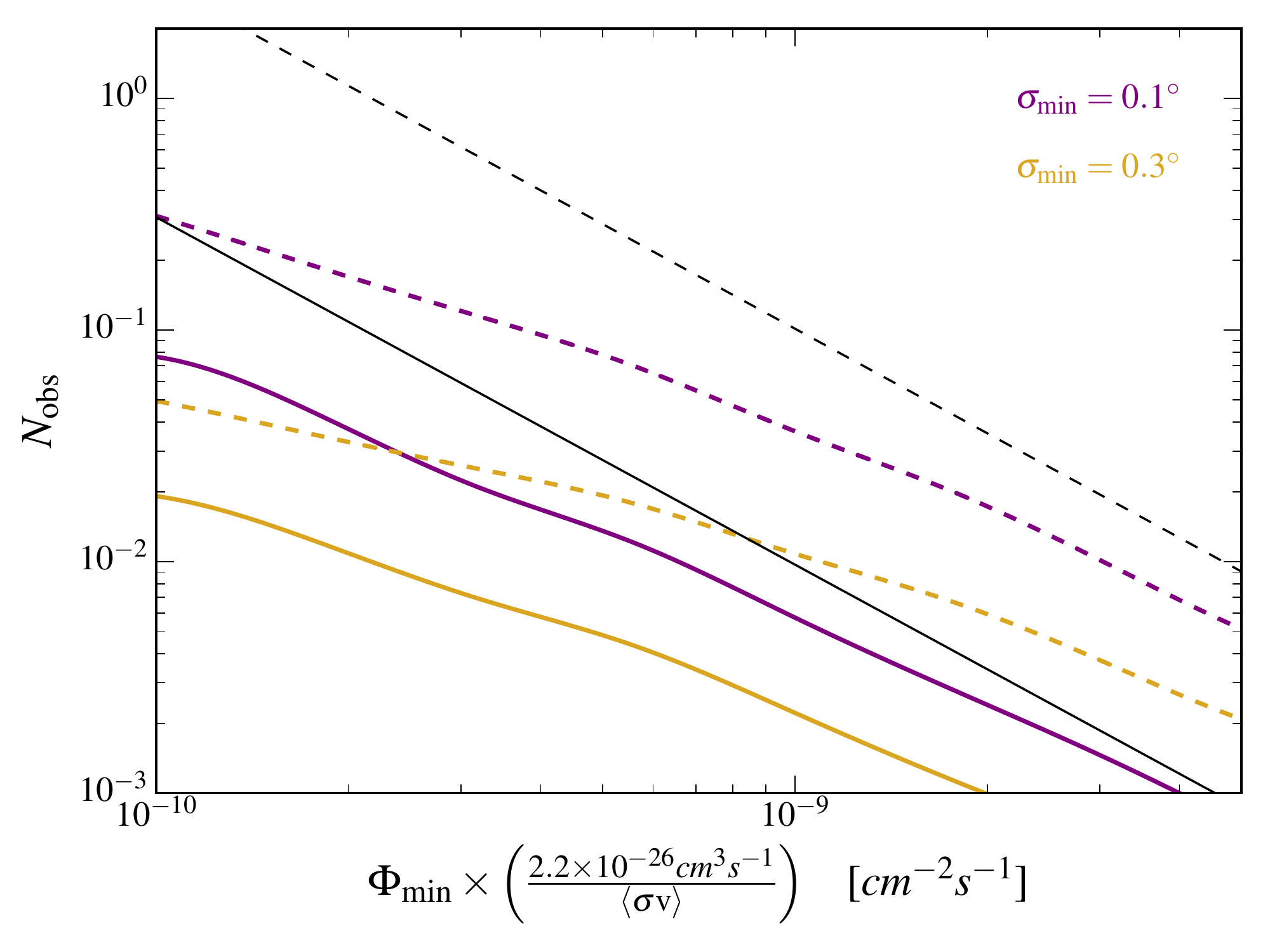}
\caption{\label{fig:extension2} Left: The fraction of subhalos with spatial extension greater than $0.1^\circ$ (purple) and $0.3^\circ$ (yellow) as a function of minimum gamma-ray flux (above 1 GeV) and annihilation cross section. Results are shown for a dark matter particle with a mass of $100$ GeV (solid) and $10$ GeV (dashed), and for the case of annihilations to $b\bar{b}$. Right: Total number of observable subhalos as a function of minimum gamma-ray flux (above 1 GeV) and annihilation cross section, for $\sigma_{\rm min} = 0$ (\ie point-like and extended, thin black), $\sigma_{\rm min} = 0.1^\circ$ (purple), $\sigma_{\rm min} = 0.3$ (yellow). As in the left panel, results are shown for a dark matter particle with a mass of $100$ GeV (solid) and $10$ GeV (dashed), and for the case of annihilations to $b\bar{b}$. }
\end{figure*}

In Fig.~\ref{fig:extension} we plot $\sigma_{_{68}}$ as a function of the distance to a given subhalo, for five values of the subhalo mass and assuming a density profile as described by Eq.~\ref{eq:den_trunc} (with $R_b$ and $\gamma$ set to their median values). This illustrates that in order for an observable subhalo to have potentially discernible extension ($\sigma_{_{68}} \gsim 0.1^{\circ}$), it must be very massive, very nearby, or both.

In the left panel of Fig.~\ref{fig:extension2}, we plot the fraction of subhalos for which $\sigma_{_{68}} > 0.1^{\circ}$ (purple) or $\sigma_{_{68}} > 0.3^{\circ}$ (yellow), as a function of the minimum gamma-ray flux and annihilation cross section. Results are shown for dark matter with a mass of 100 GeV (solid) or 10 GeV (dashed), and for the representative case of annihilations to $b\bar{b}$. The right panel of Fig.~\ref{fig:extension2} shows the total number of observable subhalos for these same candidates and minimum $\sigma_{_{68}}$ values, and compares this result with the total number of predicted subhalos (shown in black). For dark matter particles in this mass range and with an annihilation cross section of $\langle \sigma v \rangle = 2.2 \times 10^{-26}$ cm$^3/$s, we predict that approximately 10-20\% of subhalos with a gamma-ray flux above $10^{-9}$ cm$^{-2}$ s$^{-1}$ will be extended at a level of $\sigma_{_{68}}>0.3^{\circ}$ and that 40-55\% will be extended at $\sigma_{_{68}} > 0.1^{\circ}$. This can be directly compared to the degree of extension observed among those subhalo candidate sources observed by Fermi.

A recent analysis of the 12 brightest ($\Phi_{\gamma} > 10^{-9} \text{cm}^{-2}\text{s}^{-1}$) dark matter subhalo candidates in the 3FGL catalog found that three of these sources prefer a spatially extended morphology at a level of $2\Delta \ln \mathcal{L} > 4$, corresponding to $\gsim 2 \sigma$ significance~\cite{Bertoni:2016hoh}. These three sources (3FGL J2212.5+0703, 3FGL J1119.9-2204, and 3FGL J0318.1+0252) were found to be best-fit by extensions of $\sigma_{_{68}} = 0.25^{\circ}$, $0.07^{\circ}$ and $0.15^{\circ}$, respectively. The other nine sources in this sample showed little or no preference for spatial extension. Given the upper limits placed on the spatial extension of these twelve sources, eleven require $\sigma_{_{68}} < 0.3^{\circ}$ while seven require $\sigma_{_{68}} < 0.1^{\circ}$ (at the 95\% confidence level). While this manuscript was being considered for publication, Ref.~\cite{Xia:2016uog} identified an additional unassociated gamma-ray source with $\simeq 5 \sigma$ preference for a spatial extension of $\sigma \simeq 0.1^\circ$. This is particularly interesting in light of the fact that the estimated background from overlapping point sources is $\mathcal{O}(2\%)$ per source. Assessing the consistency of subhalo interpretations of these sources will be of interest in the future as the uncertainties entering subhalo analyses are further reduced.



\begin{figure*}
\center
\includegraphics[width=.7\textwidth]{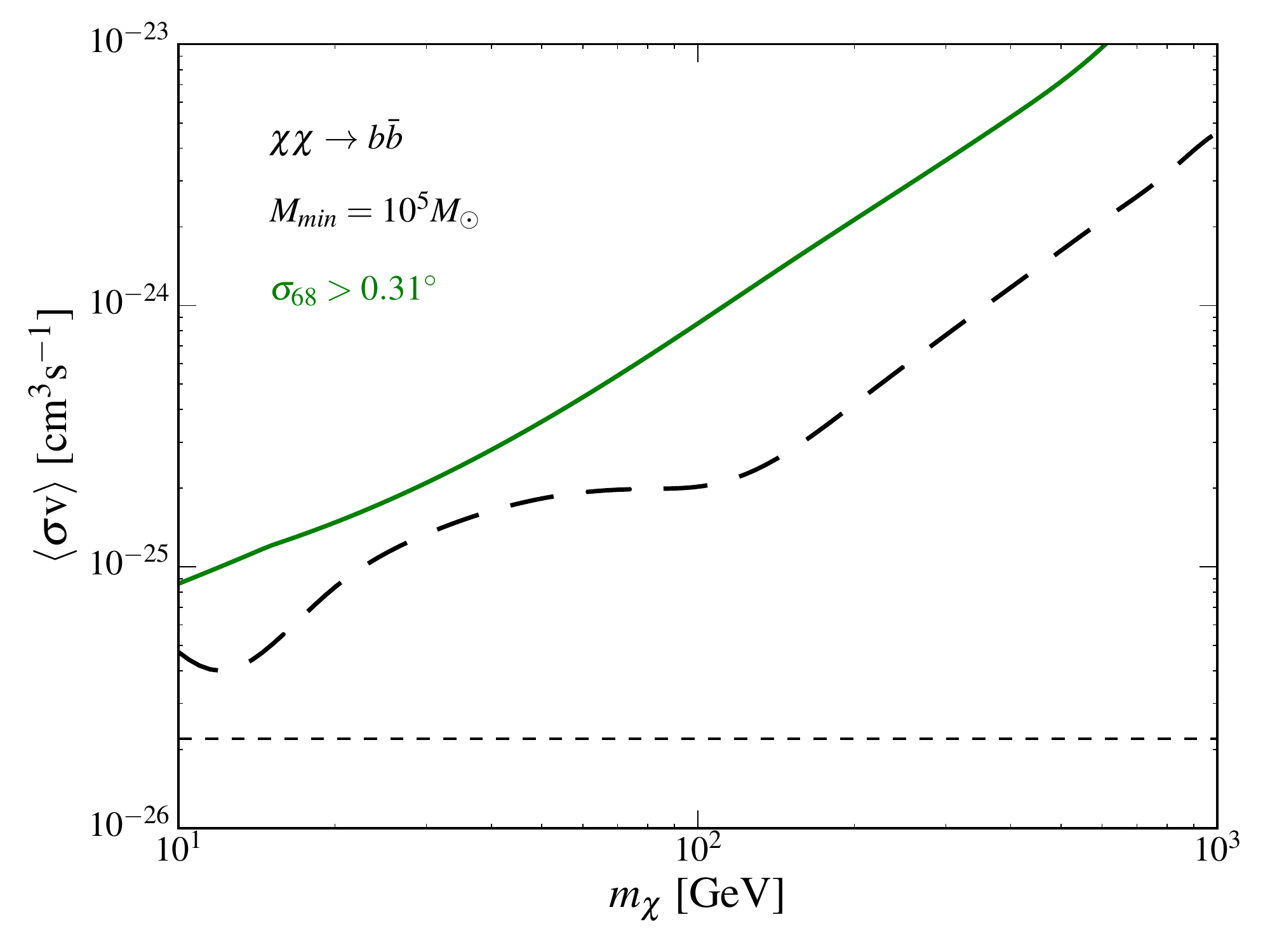}
\caption{\label{fig:extendlimits} 
The $95\%$ upper limit on the dark matter annihilation cross section for annihilation into $b\bar{b}$ derived from the non-observation of extended gamma ray sources with a flux above $10^{-9} \text{cm}^{-2}\text{s}^{-1}$ and a spatial extension $\sigma_{68}$ greater than $0.31^\circ$ (solid green). Shown for comparison are the limits derived from the total number of subhalo candidate sources as depicted in right panel of Fig.\ref{fig:limall} (dashed black).}
\end{figure*}

According to the analysis of Ref.~\cite{Bertoni:2016hoh}, none of Fermi's subhalo candidates are compatible with extension greater than $\sigma_{_{68}} > 0.31^{\circ}$. We can use this fact, in conjunction with the predicted distribution of subhalo extensions, to place an upper limit on the dark matter annihilation cross section. In Fig.~\ref{fig:extendlimits}, we plot the $95\%$ upper limit derived from the non-observation of sources with spatial extension $\sigma_{_{68}}$ greater than $0.31^\circ$ (green), for the case of annihilations to $b\bar{b}$ and a minimum subhalo mass of $10^5 M_\odot$. For comparison, we also show in this figure the limit derived from point-like sources (long dashed), assuming the same annihilation channel and minimum subhalo mass. The limit derived from the non-observation of spatially extended sources (with $\sigma_{_{68}} > 0.31^{\circ}$) is somewhat weaker than that based on the total number of sources observed. That being said, as Fermi and other gamma-ray telescopes continue to accumulate catalogs of dark matter subhalo candidate sources, spatial extension will be essential for distinguishing any subhalo population from other gamma-ray source classes.



\subsection{Uncertainties \label{sec:uncert}}

Thus far in this study, we have not addressed the many uncertainties involved in our calculations. In this section, we will discuss the most important of these uncertainties and their likely impact on our results and conclusions.

We begin by considering the density profiles of the local subhalo population. With an ideal suite of numerical simulations, one could fully resolve the profiles of individual subhalos over a wide range of scales and masses. Current simulations, however, lack the resolution to probe the inner regions such subhalos, making it difficult to distinguish between different functional forms that might describe the distributions of dark matter in these systems. We also note that current simulations are not able to resolve any small-scale structure that may be present {\it within} a given subhalo, potentially inducing a boost factor to the annihilation rate in a given subhalo. Throughout this analysis, we have conservatively chosen to neglect any boost factors to the annihilation rate.

\begin{figure*}
\center
\includegraphics[width=0.7\textwidth]{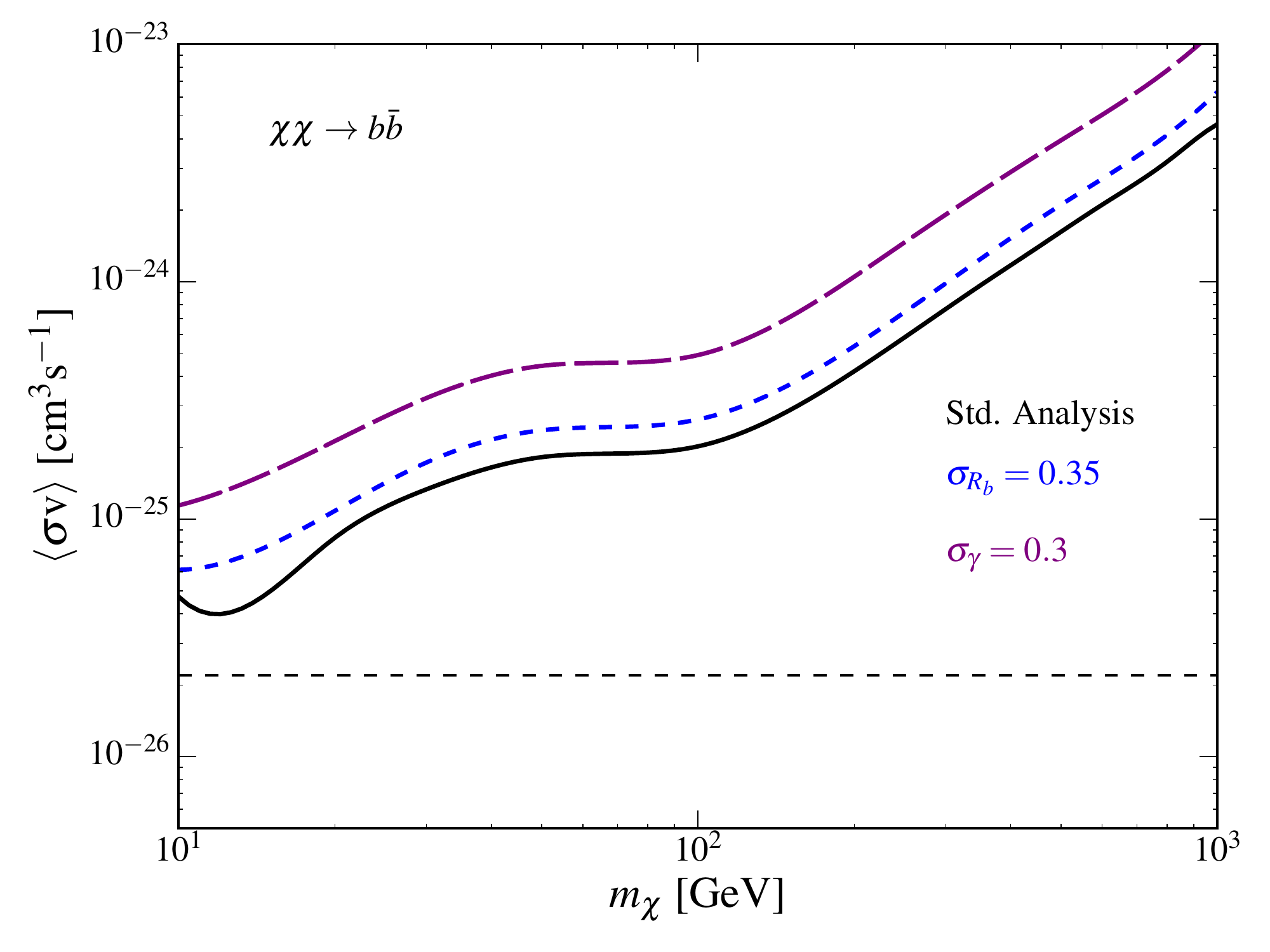}
\caption{\label{fig:limdispersion} 
The $95\%$ confidence level upper limit derived on the cross section for dark matter annihilating to $b\bar{b}$, varying independently the variance in $R_b$ (short dashed, blue) and $\gamma$ (long dashed, purple). Results are compared with the standard analysis (black). Calculations assume a minimum subhalo mass of $10^5 M_\odot$.}
\end{figure*}

Arguably, the most significant assumption we have made in our analysis is that the distributions of the parameters $\gamma$ and $R_b$ which describe the local subhalo population can be safely extrapolated from the distributions describing the subhalos located throughout the larger volume of the host halo. While the distributions of the simulated subhalos do appear to present a clear trend with respect to subhalo location with the host halo, there are simply not enough simulated subhalos in the inner tens of kiloparsecs to extract these parameters and distributions without relying on such an extrapolation. Despite the fact that it is difficult to meaningfully assess the uncertainty associated with our extrapolations of the distributions in $\gamma$ and $R_b$, it is important to understand the impact of halo-to-halo variations on predictions for the observability of subhalos. To address this question, we plot in \Fig{fig:limdispersion} the limits that would be derived should the value of $\sigma$ characterizing of the distribution in $\gamma$ (purple) and $R_b$ (blue) be reduced by a factor of $\sqrt{2}$, assuming annihilations to $b\bar{b}$ and a minimum subhalo mass of $10^5 M_\odot$. We emphasize that there is no meaningful justification for the assumed values of $\sigma$ shown in \Fig{fig:limdispersion}, but rather have included this figure to better understand how decreasing halo-to-halo variations can alter the derived limits. We believe that a proper understanding of these variations for the local population is instrumental for making concrete predictions of the observability of dark matter subhalos. Ideally, as the statistics associated with such simulations continue to improve, we hope to eventually be able to rely exclusively on simulated subhalos located in the inner regions of their host halo, eliminating the need for extrapolations in these distributions and leading to a more stable understanding of dark matter subhalos. 

\begin{figure*}
\center
\includegraphics[width=0.7\textwidth]{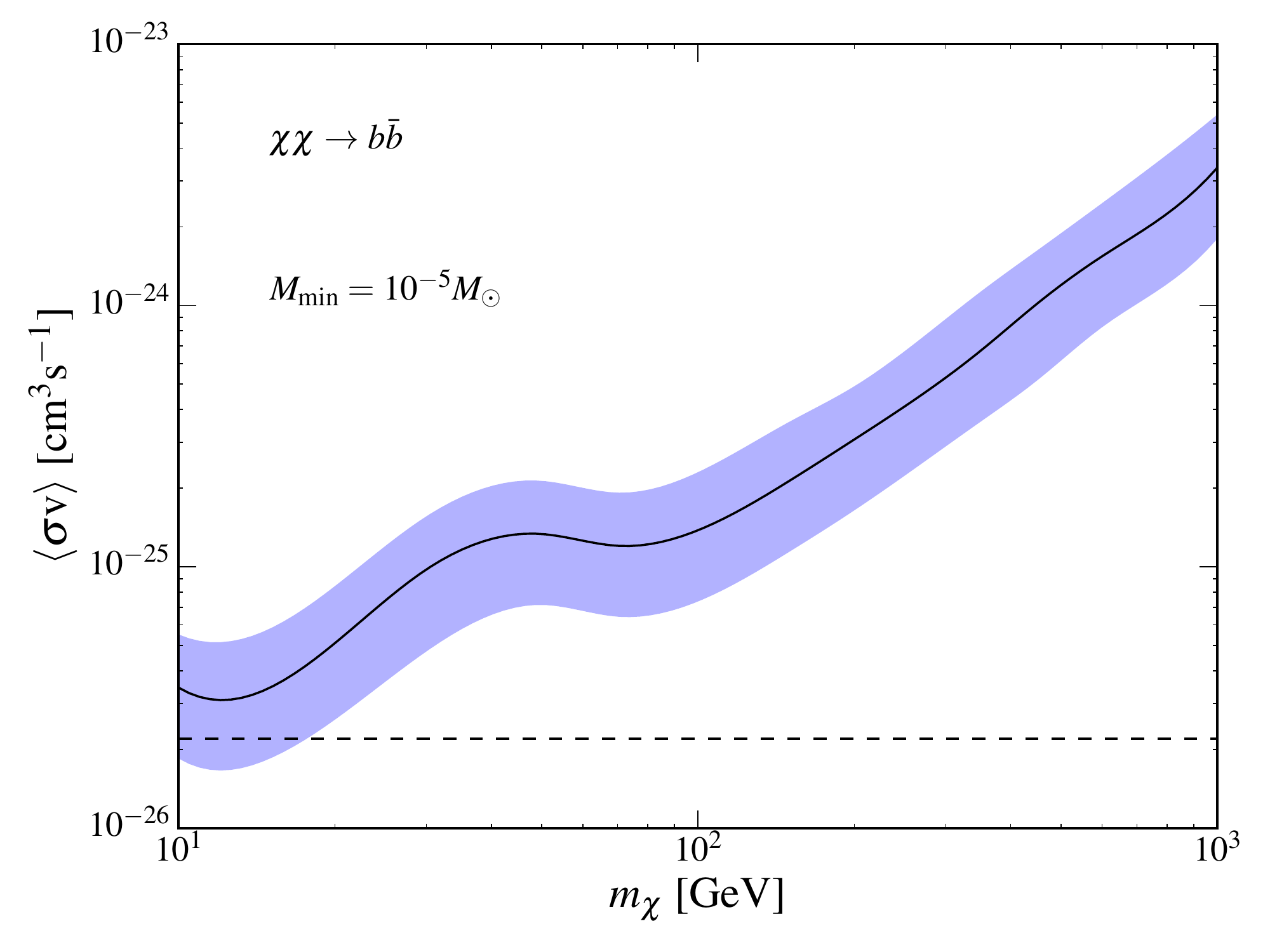}
\caption{\label{fig:uncertain} 
The $95\%$ confidence level upper limit derived on the cross section for dark matter annihilating to $b\bar{b}$, varying the exponent of the subhalo mass distribution $dN/dMdV \propto M^{\beta}$ (see Eq.~\ref{eq:numdens}) between $\beta=-1.8$ (upper boundary) and $\beta = -2.0$ (lower boundary), and adopting a minimum subhalo mass of $10^{-5} M_\odot$. The solid black contour represents the limit for our default value of $\beta =-1.9$.}
\end{figure*}

Similar to how current simulations tell us very little about the small scale structure of dark matter halos and subhalos, they are also not generally capable of revolving the lowest mass subhalos. Below roughly $10^6$ to $10^{8}\,M_{\odot}$, we are forced to extrapolate the characteristics of the local subhalo population, both in terms of the number density and mass distribution of such subhalos (see the left frame of Fig.~\ref{fig:subnumdense}), and in terms of the distributions of the halo parameters $\gamma$ and $R_b$ (see Fig.~\ref{fig:fits}). Given that the subhalo distribution extends to masses as low as $\sim$$10^{-8}$ to $10^{-3} \, M_{\odot}$ for typical WIMPs~\cite{Profumo:2006bv,Bringmann:2009vf,Cornell:2013rza,Ishiyama:2014uoa}, even modest departures from this extrapolation can have a non-negligible impact on the predicted number of observable subhalos. Some simulations actually suggest that the density profiles of the smallest scale subhalos may actually have much steeper inner slopes (with $\gamma \simeq 1.3 - 1.5$), potentially making our extrapolations slightly conservative~\cite{Ishiyama:2010es,Ishiyama:2014uoa}. To assess the uncertainty associated with the distribution of subhalos, we plot in Fig.~\ref{fig:uncertain} the upper limit on the dark matter annihilation cross section when we change the power-law slope of the subhalo mass distribution over the range of -1.8 to -2.0 (in our earlier calculations, we adopted a value of -1.9; see Eq.~\ref{eq:numdens})~\cite{Diemand:2006ik, Diemand:2008in, Garrison-Kimmel:2013eoa, Sawala:2016tlo}. Here, we have adopted a minimum subhalo mass of $10^{-5} \, M_{\odot}$ and again have considered the representative case of annihilations to $b\bar{b}$. This range of limits can vary by a factor of up to $\sim$2 (in either direction) from those presented in the left panel of Fig.~\ref{fig:limall}.

Finally, we would like to emphasize that our results are based on the subhalo populations generated in dark matter-only simulations. As the physical effects of baryons in the Milky Way are not captured in such simulations, our results do not take into account the gravitational potential of our Galaxy's stars, gas and dust. In recent years, there has been considerable progress in understanding the impact of baryons on the evolution of dark matter structure in Milky Way-like systems. In particular, some groups have attempted to capture the effect of the baryonic disk on the evolution of structure in the Milky Way without resorting to a full hydrodynamical treatment, but instead by artificially introducing a disk potential. Some of these simulations (utilizing either an artificial disk potential or a fully hydrodynamical approach) have shown that the presence of such a disk may non-negligibly reduce the local density of subhalos. For example, Refs.~\cite{Sawala:2016tlo} and~\cite{Errani2016} each find that baryonic effects reduce the local number density of subhalos by a factor of approximately $\sim$2 (see also e.g.~\cite{Despali:2016meh,Zhu:2015jwa,Calore:2016ogv}). Depending on how these baryonic effects impact the shape of the surviving subhalo density profiles, they could have a wide range of possible impacts on the resulting cross section constraints. Until such hydrodyamical effects are able to be reliably implemented with higher resolution, it will be difficult to assess their impact on the observability of the nearby dark matter subhalo population.

While this manuscript was being prepared for journal publication, a study attempting to address the baryonic impact of subhalo detectability was released~\cite{Calore:2016ogv}. Ref.~\cite{Calore:2016ogv} found a minimal impact on subhalo detectability between hydrodynamical and dark matter only simulations. We do note, however, that the conclusions of \cite{Calore:2016ogv} using the results of dark matter-only simulations differ slightly from the results shown here\footnote{Ref.~\cite{Calore:2016ogv} bases their dark matter-only results on the $\sim1200$ subhalos identified in the AQ08 simulation~\cite{Springel:2008cc}, and adopts an analysis comparable to that of~\cite{Schoonenberg:2016aml}.}. We attribute this difference primarily to the adopted subhalo parameterization.

Taking the impact of these various uncertainties together, we conclude that the predicted number of observable subhalos could quite plausibly vary from those values presented here by a factor of a few in either direction. Only with improvements in the resolution of numerical simulations (both dark matter-only and hydrodynamical simulations) will such predictions be able to be placed on firmer footing, allowing one to establish more robust limits on the dark matter annihilation cross section.

\section{Summary and Conclusion \label{sec:conc}}

In this paper, we have revisited constraints on the dark matter annihilation cross section derived from searches for dark matter subhalo candidates among Fermi's list of unassociated gamma-ray sources. We have based our calculations on the properties of over $30,000$ subhalos identified within the Via Lactea-II and ELVIS simulations, which we used to constrain the density profiles and the mass distribution of the local subhalo population. 

The density profiles of subhalos located within the innermost tens of kiloparsecs of a given host halo are significantly altered as a result of tidal stripping, and in most cases cannot be described by a traditional NFW profile. Instead, we find that these subhalos are well characterized by a power-law profile with an exponential cutoff. While the inner slope of these profiles is largely independent of the subhalo mass (consistently featuring a median value of $\langle \gamma \rangle \simeq 0.74$), the median cutoff radius is a function of mass. Using simulated subhalos from the ELVIS and Via Lactea-II simulations, we fit the distributions of these parameters as a function of the subhalo's mass and distance to the center of the host halo. From this information, we deduce the characteristics of the local subhalo population and calculate the dark matter annihilation rate within this collection of objects, determining the number of subhalos that will appear to Fermi as bright gamma-ray sources.

The limits on the dark matter annihilation cross section that have been derived in this study are somewhat weaker (by a factor of $\sim$2-3) than those presented previously by Bertoni, Hooper and Linden~\cite{Bertoni:2015mla}, and somewhat stronger than those later presented by Schoonenberg {\it et al}.~\cite{Schoonenberg:2016aml}.

We have also calculated the fraction of gamma-ray bright subhalos that are predicted to have discernible spatial extension to a telescope such as Fermi. Such information provides an important test, enabling us to potentially distinguish a dark matter subhalo from a point-like astrophysical source, such as a radio-faint pulsar. We find that for typical WIMPs, roughly 10-50\% of gamma-ray bright subhalos will be discernibly extended to Fermi, depending on the value of the dark matter's mass and annihilation cross section. This is particularly interesting in light of recent evidence for spatial extension among several of Fermi's subhalo candidates~\cite{Bertoni:2016hoh,Xia:2016uog}. The results presented here are compatible with the possibility that a significant fraction of these candidate sources could, in fact, be dark matter subhalos.

Although the limits on the dark matter annihilation cross section derived in this study are somewhat weaker than those based on observations of dwarf galaxies, the Galactic Center, and the isotropic gamma-ray background, these strategies are each subject to different uncertainties and limitations, and are thus highly complementary. Furthermore, the future prospects for dark matter subhalo searches using gamma-ray telescopes are particularly promising. In addition to further data that will be collected by Fermi, future space-based gamma-ray missions such as ComPair~\cite{Moiseev:2015lva} and e-ASTROGAM~\cite{Tatischeff:2016ykb} are anticipated to significantly improve upon the current point sensitivity at energies below $\sim$1 GeV, likely leading to the discovery of many new sources, and to the improved characterization of the sources already detected by Fermi.

\bigskip

\textbf{Acknowledgments.} We would like to thank Andrey Kravtsov for very helpful discussions. DH is supported by the US Department of Energy under contract DE-FG02-13ER41958. SW is supported under the University Research Association (URA) Visiting Scholars Award Program, and by a UCLA Dissertation Year Fellowship. Fermilab is operated by Fermi Research Alliance, LLC, under Contract No. DE-AC02-07CH11359 with the US Department of Energy.

\bibliographystyle{JHEP}
\bibliography{subhalos2016}

\end{document}